\documentclass[a4paper,12pt]{article}

\pdfoutput=1
\usepackage{jheppub}
\usepackage[T1]{fontenc}

\usepackage{graphicx}
\usepackage{cancel}
\usepackage{amssymb}
\usepackage{textcomp}
\usepackage{amsmath}
\usepackage{bm}
\usepackage{times}
\usepackage{epsfig}
\usepackage{epstopdf}
\usepackage{color}
\usepackage{multirow}

\def\lsim{\mathrel{\raise.3ex\hbox{$<$\kern-.75em\lower1ex\hbox{$\sim$}}}}
\def\gsim{\mathrel{\raise.3ex\hbox{$>$\kern-.75em\lower1ex\hbox{$\sim$}}}}

\def\beq{\begin{equation}}
\def\eeq{\end{equation}}
\def\be{\begin{equation}}
\def\ee{\end{equation}}
\def\bea{\begin{eqnarray}}
\def\eea{\end{eqnarray}}

\def\to{\rightarrow}

\title{Revisiting simplified dark matter models in terms of AMS-02 and Fermi-LAT}

\author{Tong Li}

\emailAdd{tong.li@monash.edu}
\affiliation{
ARC Centre of Excellence for Particle Physics at the Tera-scale, School of Physics and Astronomy, Monash University, Melbourne, Victoria 3800, Australia}

\abstract{
We perform an analysis of the simplified dark matter models in the light of cosmic ray observables by AMS-02 and Fermi-LAT.
We assume fermion, scalar or vector dark matter particle with a leptophobic spin-0 mediator that couples only to Standard Model quarks and dark matter via scalar and/or pseudo-scalar bilinear. The propagation and injection parameters of cosmic rays are determined by the observed fluxes of nuclei from AMS-02. We find that the AMS-02 observations are consistent with the dark matter framework within the uncertainties. The AMS-02 antiproton data prefer 30 (50) GeV - 5 TeV dark matter mass and require an effective annihilation cross section in the region of $4 \times 10^{-27} \ (7 \times 10^{-27})$ -- $4 \times 10^{-24}$ ${\rm cm}^3/{\rm s}$ for the simplified fermion (scalar and vector) dark matter models. The cross sections below $2 \times 10^{-26} \ {\rm cm}^3/{\rm s}$ can evade the constraint from Fermi-LAT dwarf galaxies for about 100 GeV dark matter mass.
}

\begin{document}

\maketitle
\flushbottom
\newpage

\section{Introduction}

The measurements of Galactic cosmic rays provide crucial information to understand their own source and propagation and further confine the possibly new fundamental particle physics such as dark matter (DM) annihilation. Cosmic rays in the Galaxy are categorized into primary and secondary types according to their different origins~\cite{1964ocr..book.....G, Blandford:1987pw, Stawarz:2009ig, Aharonian:2011da}. Primary cosmic rays are those which are created by astrophysical sources, while their initial spectrum and composition change and thus emit secondary cosmic rays as a result of interacting with matter in the interstellar medium (ISM).
The secondary-to-primary ratio of cosmic ray nuclei (such as the Boron-to-Carbon ratio B/C) and the ratio of secondary isotopes (such as the Beryllium ratio $\rm ^{10}Be/^{9}Be$) are widely employed to constrain the cosmic ray propagation parameters as they are respectively sensitive to the traveling path and the lifetime of cosmic rays in the Galaxy.
The source injection parameters of cosmic ray nuclei can be constrained by the measured proton flux data.

Recently, AMS-02 collaboration released abundant and precise data on the cosmic ray nuclei, e.g. proton~\cite{Aguilar:2015ooa}, B/C~\cite{Aguilar:2016vqr}, etc.
Combining with old data sets of $\rm ^{10}Be/^{9}Be$ from ACE~\cite{ACE} and proton from PAMELA~\cite{Adriani:2013as}, one can constrain the propagation and source injection parameters in an efficiently statistical method. Based on these parameters, an up-to-date antiproton cosmic rays as the secondary production from colliding protons with ISM can be obtained with high precision. Given this astrophysical background of antiproton, we are enabled to confine extra compositions in cosmic rays such as annihilating dark matter which also produces antiproton, in the light of the antiproton flux data newly reported by AMS-02~\cite{Aguilar:2016kjl}.

In addition to charged cosmic rays, gamma ray flux is also an observable that dark matter can produce potentially. Dwarf galaxies are one of the best places with a large abundance of dark matter and thus the bright targets to search for gamma rays from dark matter annihilation. The Fermi Large Area Telescope (LAT) has looked for gamma ray emission from the dwarf spheroidal satellite galaxies (dSphs) of the Milky Way and detected no excess. Fermi-LAT thus placed upper limit on the dark matter annihilation cross section from a combined analysis of 15 Milky Way dSphs~\cite{Ackermann:2015zua} and recently updated the result with more candidate dSphs and increased sensitivity~\cite{Fermi-LAT:2016uux}. They are generally most stringent constraints for dark matter annihilating into quark or gluon channels~\cite{Elor:2015bho}.

In this work, we examine the constraints set by the AMS-02 antiproton data and the Fermi-LAT dSphs on the simplified models with weakly interacting mass particle (WIMP) as dark matter. Specifically we consider fermion, scalar and vector dark matter denoted by $\chi$, $\phi$ and $X$ respectively, with mediators that only couple to the Standard Model (SM) quarks and dark matter particles. This leptophobic framework is widely used to analyze the data for dark matter search in indirect detection (ID) and direct detection (DD) and collider experiments~\cite{Buchmueller:2013dya,Arina:2014yna,Alves:2015pea,Abdallah:2015ter,Boveia:2016mrp,Arina:2016cqj,Ismail:2016tod,Balazs:2017hxh}.
It uses minimal and general theoretical assumptions with only two parameters, i.e. the dark matter mass and the mediator mass.

In the simplified framework the annihilation of dark matter in s-channel occurs through the exchange of either a spin-0 or spin-1 mediator. The general interactions between the mediator and dark matter or SM quarks are the Lorentz-invariant combinations of the following bilinears
\begin{eqnarray}
&&\bar{\chi}\chi, \bar{\chi}\gamma^5\chi, \bar{\chi}\gamma^\mu\chi \ ({\rm Dirac \ only}), \bar{\chi}\gamma^\mu\gamma^5\chi; \phi^\dagger \phi, \phi^\dagger \overleftrightarrow{\partial_\mu}\phi \ ({\rm complex \ only}); X_\mu^\dagger X^\mu, X^\dagger_\mu \partial_\nu X^\nu ; \nonumber \\
&&m_q\bar{q}q, m_q\bar{q}\gamma^5q, \bar{q}\gamma^\mu q, \bar{q}\gamma^\mu\gamma^5 q.\nonumber
\end{eqnarray}
As dictated by minimal flavor violation, the couplings of scalar and pseudo-scalar quark bilinears are scaled by quark mass $m_q$ and those of vector and axial-vector bilinears are chosen to be universal~\cite{Buras:2000dm,Goodman:2010ku}. The spin-1 mediator scenario via vector or axial-vector interaction is thus highly constrained by the dijet limit for $Z'$ search at the Large Hadron Collider (LHC)~\cite{Khachatryan:2016ecr,Sirunyan:2016iap,StevenSchramm}. The collider search is yet less sensitive to the detection of spin-0 mediator scenario through scalar or pseudo-scalar interaction as a result of the $m_q^2$ suppression coming from the Yukawa coupling. Among the structure combinations in the spin-0 mediator scenario only four forms lead to annihilation cross section without velocity suppression, denoted by $D_2$, $D_4$, $C$ and $V$ using the notation of effective field theories (EFTs)~\cite{Goodman:2010ku}, as shown in Table~\ref{tab:operator}. Moreover, for these models, the nucleon-WIMP scattering rates are either suppressed by the spin of the target nucleus and/or dark matter and the scattering momentum exchange~\cite{DeSimone:2016fbz}, rendering weak DD constraints. We thus investigate the sensitivity of AMS-02 and Fermi-LAT indirect observables to the detection of simplified dark matter models $D_2$, $D_4$, $C$ and $V$.

\begin{table}[h]
\begin{center}
\begin{tabular}{|c|c|c|c|}
        \hline
        Interations & ID & DD & Collider\\
        \hline
        ${\rm D_2}: \bar{\chi}\gamma_5\chi\oplus\bar{q}q$ & 1 & $\mathbf{s}_\chi\cdot \mathbf{q}$ & $m_q^2$ \\
        \hline
        ${\rm D_4}: \bar{\chi}\gamma_5\chi\oplus\bar{q}\gamma_5 q$ & 1 & $(\mathbf{s}_\chi\cdot \mathbf{q})(\mathbf{s}_N\cdot \mathbf{q})$ & $m_q^2$\\
        \hline
        ${\rm C}: \phi^\dagger\phi\oplus\bar{q}\gamma_5 q$ & 1 & $\mathbf{s}_N\cdot \mathbf{q}$ & $m_q^2$ \\
        \hline
        ${\rm V}: X^\dagger_\mu X^\mu\oplus\bar{q}\gamma_5 q$ & 1 & $\mathbf{s}_N\cdot \mathbf{q}$ & $m_q^2$\\
        \hline
\end{tabular}
\end{center}
\caption{Interactions considered in this work and their suppression effects for ID, DD and collider search. $\mathbf{s}_\chi$ ($\mathbf{s}_N$) is the spin of dark matter (the target nucleus) and $\mathbf{q}$ is the scattering momentum exchange.}
\label{tab:operator}
\end{table}

This paper is organized as follows. In Sec.~\ref{sec:Models}, we discuss the simplified dark matter models we use. In Sec.~\ref{sec:Observable} we describe the observales from AMS-02 and Fermi-LAT. Our numerical results are given in Sec.~\ref{sec:Results}.
Finally, in Sec.~\ref{sec:Concl} we summarize our conclusions.

\section{The Simplified Dark Matter Models}
\label{sec:Models}

In this section, we describe the considered simplified dark matter models in Table~\ref{tab:operator}.
The dark matter particles ($\chi,\phi,X$) couple to the SM quarks through a spin-0 mediator $S_2$, $S_4$, $S_C$ or $S_V$ corresponding to structure $D_2$, $D_4$, $C$ or $V$ respectively. The corresponding interactions are as follows~\cite{Berlin:2014tja,Abdallah:2015ter}
\begin{eqnarray}
{\cal L}_{\rm D2} &=& -ig_{\chi}^{\rm D2} S_2\bar{\chi}\gamma_5 \chi - S_2\sum_{q=u,d,s,c,b,t}g_q^{\rm D2} {m_q\over v_0}\bar{q} q,
\label{eq:interactionD2}\\
{\cal L}_{\rm D4} &=& -ig_{\chi}^{\rm D4} S_4\bar{\chi}\gamma_5 \chi - iS_4\sum_{q=u,d,s,c,b,t}g_q^{\rm D4} {m_q\over v_0}\bar{q}\gamma_5 q,
\label{eq:interactionD4}\\
{\cal L}_{\rm C} &=& -g_{\phi}^{\rm C} m_\phi S_C \phi^\dagger\phi - iS_C\sum_{q=u,d,s,c,b,t}g_q^{\rm C} {m_q\over v_0}\bar{q}\gamma_5 q,
\label{eq:interactionC}\\
{\cal L}_{\rm V} &=& -g_{X}^{\rm V} m_X S_V X^\dagger_\mu X^\mu - iS_V\sum_{q=u,d,s,c,b,t}g_q^{\rm V} {m_q\over v_0}\bar{q}\gamma_5 q,
\label{eq:interactionV}
\end{eqnarray}
where $v_0=246$ GeV is the Higgs vacuum expectation value. Following the general choices in the analysis of dark matter searches in literatures, we take $g_{\chi}^{\rm D2}=g_{\chi}^{\rm D4}=g_q^{\rm D2}=g_q^{\rm D4}=g_{\phi}^{\rm C}=g_{X}^{\rm V}=g_q^{\rm C}=g_q^{\rm V}=1$ in the calculations below. Under the above assumptions the dark matter models are described by two parameters, i.e. the dark matter mass $m_{\rm DM}=m_\chi, m_\phi, m_X$ and the mediator mass $m_{\rm Med}=m_{S_2}$, $m_{S_4}$, $m_{S_C}$ or $m_{S_V}$. We scan these parameters in the following range
\begin{eqnarray}
5 \ {\rm GeV} < m_{\rm DM}, m_{\rm Med} < 10 \ {\rm TeV}.
\end{eqnarray}

Given the interactions in Eqs.~(\ref{eq:interactionD2}) and (\ref{eq:interactionD4}), the pairs of dark matter particle can either annihilate into SM quark or gluon pairs via ${\rm DM} \ {\rm DM} \to {\rm Med}\to \bar{q}q, gg$ (${\rm DM}=\chi,\phi,X$; ${\rm Med}=S_2,S_4,S_C,S_V$)
or annihilate into four SM quarks or gluons via two mediators in t-channel ${\rm DM} \ {\rm DM} \to {\rm Med} \ {\rm Med} \to 4 \ {\rm quarks}, 4 \ {\rm gluons}$
when kinematically allowed. The energy distribution of cosmic rays produced in the annihilation, as a result, is the sum of 2-body spectrum and 4-body spectrum
\begin{eqnarray}
dN_i/ dE=(dN_i/ dE)_{\rm 2-body}+(dN_i/ dE)_{\rm 4-body},
\end{eqnarray}
where $i$ denotes the cosmic ray species, i.e. $\bar{p},\gamma$ here. The two types of spectrum are both the annihilation-fraction-weighted sum of the differential yields into specific final states. For different quark or gluon final states in 2-body spectrum we use PPPC4DMID~\cite{Cirelli:2010xx} to generate the differential yield which is weighted by the corresponding annihilation fraction, i.e. $\langle \sigma_{\rm ann} v\rangle_{\bar{q}q}/\langle \sigma_{\rm ann} v\rangle$ and $\langle \sigma_{\rm ann} v\rangle_{gg}/\langle \sigma_{\rm ann} v\rangle$. The 4-body cosmic ray spectrum is given by the spectrum of the mediator decay in its rest frame followed by a Lorentz boost~\cite{Elor:2015tva,Elor:2015bho}. The spectrum is then weighted by the product of annihilation fraction and decay branching ratio of the mediator, i.e. ${\langle \sigma_{\rm ann} v\rangle_{\rm Med}\over \langle \sigma_{\rm ann} v\rangle} {\Gamma_{{\rm Med}\to q\bar{q}}\over \Gamma_{\rm Med}}$ and ${\langle \sigma_{\rm ann} v\rangle_{\rm Med}\over \langle \sigma_{\rm ann} v\rangle} {\Gamma_{{\rm Med}\to gg}\over \Gamma_{\rm Med}}$, to give $(dN_i/ dE)_{\rm 4-body}$~\cite{Li:2016uph}. The expressions of mediator decay widths and dark matter annihilation cross sections are collected in Appendix.

\section{Indirect Observables from AMS-02 and Fermi-LAT}
\label{sec:Observable}
In this section we describe the observables of antiproton flux and gamma ray measured by AMS-02 and Fermi-LAT respectively.

\subsection{Antiproton flux from AMS-02}
\label{sec:Propagation}

The two key unknowns about cosmic rays in the Galaxy are their origin and propagation. The propagation of cosmic rays can be described as the process of diffusion. The diffusion process is written in the form of the transport equation below~\cite{Strong:2007nh}
\begin{eqnarray}
{\partial \psi\over \partial t}&=&Q(\vec{r},p)+\vec{\nabla}\cdot \left(D_{xx}\vec{\nabla}\psi-\vec{V}\psi\right)+{\partial\over \partial p}p^2D_{pp}{\partial\over \partial p}{1\over p^2}\psi \nonumber\\
&&-{\partial\over \partial p}\left[\dot{p}\psi-{p\over 3}\left(\vec{\nabla}\cdot \vec{V}\right)\psi\right]-{\psi\over \tau_f}-{\psi\over \tau_r},
\label{propagation}
\end{eqnarray}
where $\psi(\vec{r},t,p)$ is the density of cosmic rays, $\vec{V}$ is the convection velocity and $\tau_f (\tau_r)$ is the time scale for fragmentation (radioactive decay). $\dot{p}$ is the momentum loss rate. The convection terms in the above equation are induced by the Galactic wind. The diffusion in momentum space governs the reacceleration process. In this case the diffusion coefficient in momentum space, i.e. $D_{pp}$, is related to the spatial coefficient $D_{xx}$ and the Alfven velocity $v_A$~\cite{Ginzburg:1990sk}:
\begin{eqnarray}
D_{pp}D_{xx}={4p^2v_A^2\over 3\delta(4-\delta^2)(4-\delta)w},
\end{eqnarray}
with the level of the interstellar turbulence parameter $w$ being 1. The spatial diffusion coefficient is usually written in this form
\begin{eqnarray}
D_{xx}=\beta D_0 (R/R_0)^\delta ,
\end{eqnarray}
with $R$ and $\beta$ being the rigidity and particle velocity divided by light speed respectively. This transport equation is numerically solved based on given boundary conditions, that is, the cosmic ray density $\psi$ vanishes at the radius $R_h$ and the height $z_0$ of the cylindrical diffusion halo.

In Eq.~(\ref{propagation}), the source term can be written by the product of the spatial distribution and the injection spectrum function for cosmic ray species $i$
\begin{eqnarray}
Q_i(\vec{r},p)=f(r,z)q_i(p) .
\end{eqnarray}
We use the following supernova remnants distribution for the spatial distribution of the primary cosmic rays
\begin{eqnarray}
f(r,z)=f_0\left({r\over r_\odot}\right)^a{\rm exp}\left(-b \ {r-r_\odot\over r_\odot}\right){\rm exp}\left(-{|z|\over z_s}\right),
\label{snr}
\end{eqnarray}
where $r_\odot=8.5 \ {\rm kpc}$ is the distance between the Sun and the Galactic center, the height of the Galactic disk is $z_s=0.2 \ {\rm kpc}$. The two parameters $a$ and $b$ are chosen to be 1.25 and 3.56, respectively~\cite{Lin:2014vja}.
The following power law with one break is assumed for the injection spectrum of various nuclei
\begin{eqnarray}
q_i&\propto& \left\{
                \begin{array}{ll}
                  \left(R/R_{\rm br}^p\right)^{-\nu_1}, & R\leq R_{\rm br}^p \\
                  \left(R/R_{\rm br}^p\right)^{-\nu_2}, & R> R_{\rm br}^p
                \end{array}
              \right. \ ,
\label{injection}
\end{eqnarray}
where the rigidity break $R_{\rm br}^p$ and power law indexes $\nu_1,\nu_2$ are injection parameters.

The above propagation parameters and source injection parameters can be constrained by fitting the ratios of nuclei, i.e. the Boron-to-Carbon ratio (B/C) and the Beryllium ratio ($\rm ^{10}Be/^{9}Be$), and proton flux data respectively. There existed many attempts fitting these parameters since the release of new AMS-02 nuclei data~\cite{Cuoco:2016eej,Cui:2016ppb,Feng:2016loc,Huang:2016tfo,Lin:2016ezz,Jin:2017iwg,Yuan:2017ozr,Niu:2017qfv}. In particular Ref.~\cite{Yuan:2017ozr} combined the relevant data sets of cosmic rays measured by ACE, PAMELA and AMS-02 in their Markov Chain Monte Carlo (MCMC) sampling algorithm and gave the fitted results for different propagation models.
As shown in Table~\ref{tab:parameter}, we adopt the values of propagation/injection parameters in the diffusion reacceleration (DR) model which fits both the B/C and proton fluxes well compared with convection models and does not need additional phenomenological modification of the diffusion coefficient~\cite{Yuan:2017ozr}.
The values of Fisk potential are the approximate constants of the time-varying modulation form employed in the references~\cite{Cui:2016ppb,Yuan:2017ozr}.

\begin{table}[h]
\begin{center}
\resizebox{15cm}{!} {
\begin{tabular}{|c|c|c|c|c|c|c|c|}
        \hline
        propagation & value && nucleon injection & value && solar modulation & value\\
        \hline
        $D_0 \ (10^{28} \ {\rm cm}^2 \ {\rm s}^{-1})$ & 7.24 && $\nu_1$ & 1.69 && $\phi_{p} \ ({\rm MV})$ & 550\\
        \hline
        $\delta$ & 0.38 && $\nu_2$ & 2.37 && $\phi_{\bar{p}} \ ({\rm MV})$ & 400\\
        \hline
        $R_0 \ ({\rm GV})$ & 4 && $R_{\rm br}^p$ \ ({\rm GV}) & 12.88 && $-$ & $-$ \\
        \hline
        $v_A \ ({\rm km} \ {\rm s}^{-1})$ & 38.5 && $A_p$ (see caption) & 4.498 && $-$ & $-$\\
        \hline
        $z_0$ \ ({\rm kpc}) & 5.93 &&    $-$    & $-$ && $-$ & $-$ \\
        \hline
\end{tabular}}
\end{center}
\caption{Parameters of propagation, nucleon injection and solar modulation and their values adopted in our numerical analysis.  The proton flux is normalized to  $A_p$ at 100 GeV in the units of $10^{-9} \ {\rm cm}^{-2} \ {\rm s}^{-1} \ {\rm sr}^{-1} \ {\rm MeV}^{-1}$.}
\label{tab:parameter}
\end{table}

Compared to measured data, the benchmark model of propagation generally underproduces the antiproton cosmic ray at high energies. The dark matter annihilation can also produce antiprotons so as to compensate this discrepancy.
The dark matter source term contributing to the cosmic ray species $i$ is given by
\begin{eqnarray}
Q_i^{\rm DM}(r,p)=\frac{\rho_{\rm DM}^2(r)\langle \sigma_{\rm ann} v\rangle}{a m_{\rm DM}^2}\frac{dN_i}{dE}, \ \ \ i=\bar{p} \ \ {\rm here}
\label{dmsource}
\end{eqnarray}
where $a=2 \ (4)$ for self-conjugate (non self-conjugate) dark matter.
We use a generalized Navarro-Frenk-White (NFW) profile to describe dark matter spatial distribution~\cite{NFW}
\begin{eqnarray}
\rho_{\rm DM}(r)=\rho_0\frac{(r/r_s)^{-\gamma}}{(1+r/r_s)^{3-\gamma}}.
\end{eqnarray}
The NFW profile is a traditional benchmark choice motivated by N-body simulations.
The inner slope of the halo profile is chosen to be $\gamma=1$ and the radius of the galactic diffusion disk is $r_s=20$ kpc. The coefficient $\rho_0$ is thus set to be $0.26 \ {\rm GeV/cm^3}$ to give the local dark matter density $\rho_{\rm DM}(r_\odot)=0.3 \ {\rm GeV/cm^3}$.

As mentioned above, for each set of dark matter mass and mediator mass, we first generate the antiproton spectrum $dN_{\bar{p}}/dE$, and calculate the dark matter annihilation cross section. These dark matter model dependent variables are then passed into the public code Galprop v54~\cite{Strong:1998pw, Moskalenko:2001ya, Strong:2001fu, Moskalenko:2002yx, Ptuskin:2005ax} to ensure that near Earth cosmic ray fluxes from dark matter annihilation and background are obtained in a consistent way~\cite{Balazs:2015iwa}. The obtained cosmic ray fluxes, together with the experimental data points, are put into a composite likelihood function, defined as
\begin{eqnarray}
- 2\ln{\cal L}= \sum_i {(f_i^{\rm th}-f_i^{\rm exp})^2\over \sigma_i^2} .
\label{sum}
\end{eqnarray}
Here $f_i^{\rm th}$ are the theoretical predictions and $f_i^{\rm exp}$ are the corresponding central value of the measured data. We stipulate a 50\% uncertainty of the theoretical prediction according to the estimates in Refs.~\cite{Cui:2016ppb,Trotta:2010mx, Auchettl:2011wi, Giesen:2015ufa}. This 50\% takes into account, amongst other, the uncertainty related to the fixed propagation parameters. The theoretical and experimental uncertainties are then combined in quadrature to give the $\sigma_i$. Note that the AMS-02 analyses of antiproton flux and antiproton-to-proton ratio were based on the same antiproton events.
Although their systematic uncertainties are different, the two data sets might be correlated. In order to avoid the possible correlation, the sum in Eq.~(\ref{sum}) runs over only the AMS-02 antiproton flux data (57 points).

\subsection{Dwarf galaxy constraint from Fermi-LAT}
\label{sec:Fermi}
For individual dwarf galaxy target, Fermi-LAT tabulated the delta-log-likelihood values as a function of the energy flux bin-by-bin~\cite{Ackermann:2015zua} and newly reported an update in Ref.~\cite{Fermi-LAT:2016uux}.
The gamma ray energy flux from dark matter annihilation for $j$th energy bin is given by
\begin{eqnarray}
\Phi^E_{j,k}(m_{\rm DM},\langle \sigma_{\rm ann} v\rangle,J_k)={1\over 4\pi}\frac{\langle \sigma_{\rm ann} v\rangle}{a m_{\rm DM}^2}J_k\int^{E^{\rm max}_j}_{E^{\rm min}_j}E\frac{dN_i}{dE}dE, \ \ \ i=\gamma \ \ {\rm here}
\end{eqnarray}
where $J_k$ is the J factor for $k$th dwarf. One can see that the energy flux is only dependent on three parameters, i.e. $m_{\rm DM}$, $\langle \sigma_{\rm ann} v\rangle$ and $J_k$, and calculable for any dark matter annihilating process induced by the above simplified models. The likelihood for $k$th dwarf is
\begin{eqnarray}
\mathcal{L}_k(m_{\rm DM},\langle \sigma_{\rm ann} v\rangle,J_k)=\mathcal{L}_J(J_k|\bar{J}_k,\sigma_k)\prod_j \mathcal{L}_{j,k}(\Phi^E_{j,k}(m_{\rm DM},\langle \sigma_{\rm ann} v\rangle,J_k))
\end{eqnarray}
where $\mathcal{L}_{j,k}$ is the tabulated likelihood provided by Fermi-LAT for each dwarf and calculated energy flux and the uncertainty of the J factors is taken into account by profiling over $J_k$ in the likelihood below
\begin{eqnarray}
\mathcal{L}_J(J_k|\bar{J}_k,\sigma_k)={1\over \ln(10)J_k\sqrt{2\pi}\sigma_k}\times e^{-(\log_{10}(J_k)-\log_{10}(\bar{J}_k))^2/2\sigma_k^2}
\end{eqnarray}
with the measured J factor $\bar{J}_k$ and error $\sigma_k$. A joint likelihood for all dwarfs is then performed as
\begin{eqnarray}
\mathcal{L}(m_{\rm DM},\langle \sigma_{\rm ann} v\rangle,\mathbb{J})=\prod_k \mathcal{L}_k(m_{\rm DM},\langle \sigma_{\rm ann} v\rangle,J_k)
\end{eqnarray}
where $\mathbb{J}$ is the set of J factors $J_k$. In our implementation we adopt the corresponding values of $\mathcal{L}_{j,k}$ and $\bar{J}_k, \sigma_k$ for 19 dwarf galaxies considered in Ref.~\cite{Fermi-LAT:2016uux}. Specifically they are Bootes I, Canes Venatici I, Canes Venatici II, Carina, Coma Berenices, Draco, Fornax, Hercules, Leo I, Leo II, Leo IV, Leo V, Reticulum II, Sculptor, Segue 1, Sextans, Ursa Major I, Ursa Major II, and Ursa Minor.

As Fermi-LAT found no gamma ray excess from the dSphs, one can set upper limit on the annihilation cross section for a given $m_{\rm DM}$ by taking J factors as nuisance parameters in the maximum likelihood analysis. Following Fermi's approach, the delta-log-likelihood is given by
\begin{eqnarray}
-2\Delta \ln \mathcal{L}(m_{\rm DM},\langle \sigma_{\rm ann} v\rangle)=-2\ln\left({\mathcal{L}(m_{\rm DM},\langle \sigma_{\rm ann} v\rangle,\widehat{\widehat{\mathbb{J}}})\over \mathcal{L}(m_{\rm DM},\widehat{\langle \sigma_{\rm ann} v\rangle},\widehat{\mathbb{J}})}\right)
\end{eqnarray}
where $\widehat{\langle \sigma_{\rm ann} v\rangle}$ and $\widehat{\mathbb{J}}$ maximize the likelihood at any given $m_{\rm DM}$, and $\widehat{\widehat{\mathbb{J}}}$ maximize the likelihood for given $m_{\rm DM}$ and $\langle \sigma_{\rm ann} v\rangle$.
The 95\% C.L. upper limit on annihilation cross section for a given $m_{\rm DM}$ is determined by demanding $-2\Delta \ln\mathcal{L}(m_{\rm DM},\langle \sigma_{\rm ann} v\rangle)\leq 2.71$. We perform the likelihood analysis using Minuit~\cite{James:1975dr}. If the annihilation cross section calculated by a certain set of $m_{\rm DM}$ and the mediator mass is greater than the upper limit, we claim the corresponding set of $m_{\rm DM}, m_{\rm Med}$ is excluded by Fermi-LAT dSphs.

\section{Results}
\label{sec:Results}

As varying the two mass parameters in simplified dark matter models and thus the likelihood function in Eq.~(\ref{sum}), we can fit the AMS-02 antiproton flux data and obtain the confidence regions of dark matter model parameters. We calculate $2\sigma$ confidence region by increasing the likelihood function from its best fit value, whilst scanning the two mass parameters, until $-2\ln{\cal L}$ changes by 6.18. The dark matter contributions to the observables are then calculated using the dark matter model parameters in the $2\sigma$ confidence region. Our likelihood function does not include the antiproton-to-proton ratio data. Rather, after we extract the dark matter model parameters, the antiproton-to-proton ratio using the fitted parameters is given as a cross check.

\begin{figure}[t]
\begin{center}
\includegraphics[scale=1,width=7cm]{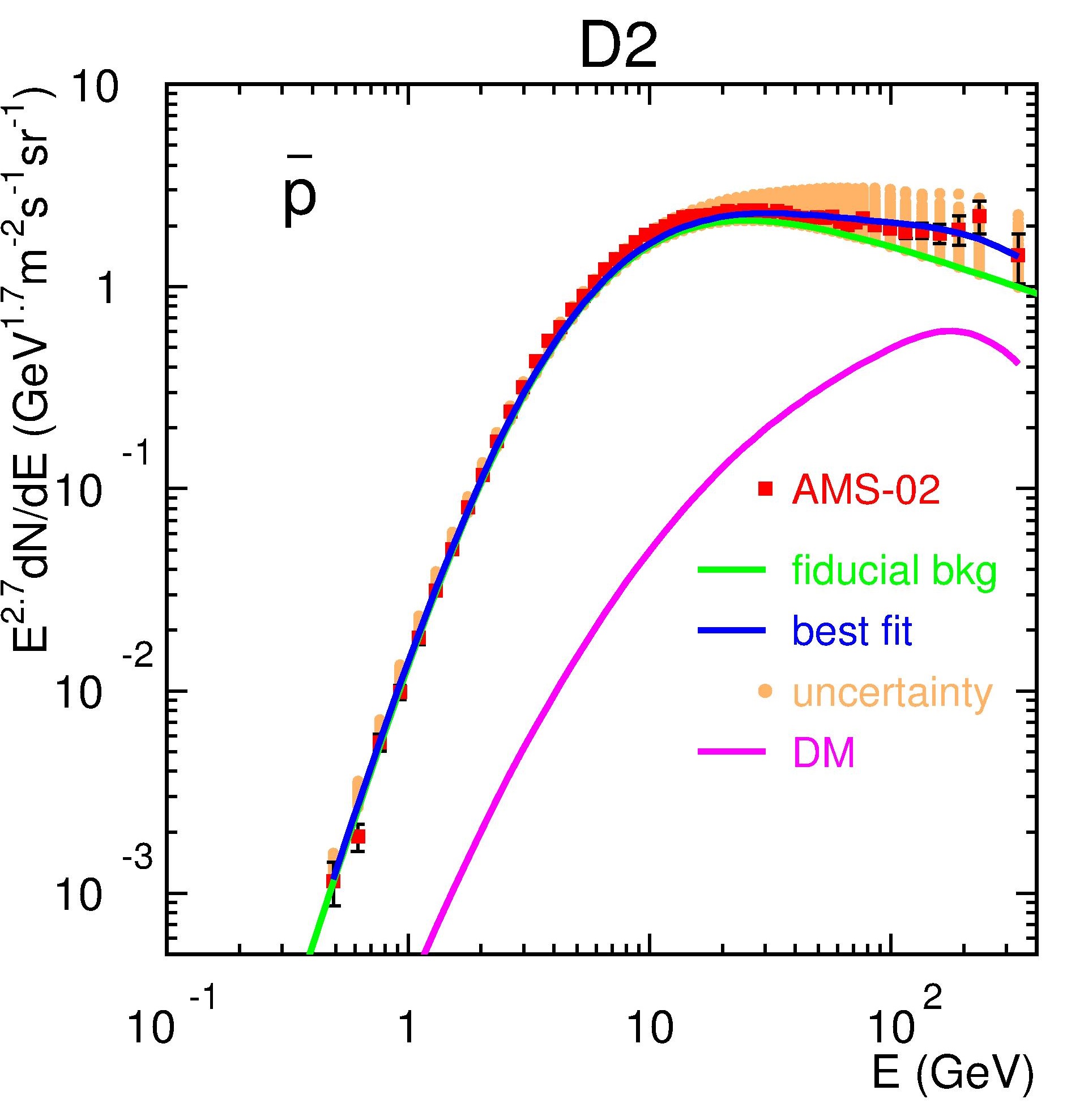}
\includegraphics[scale=1,width=7cm]{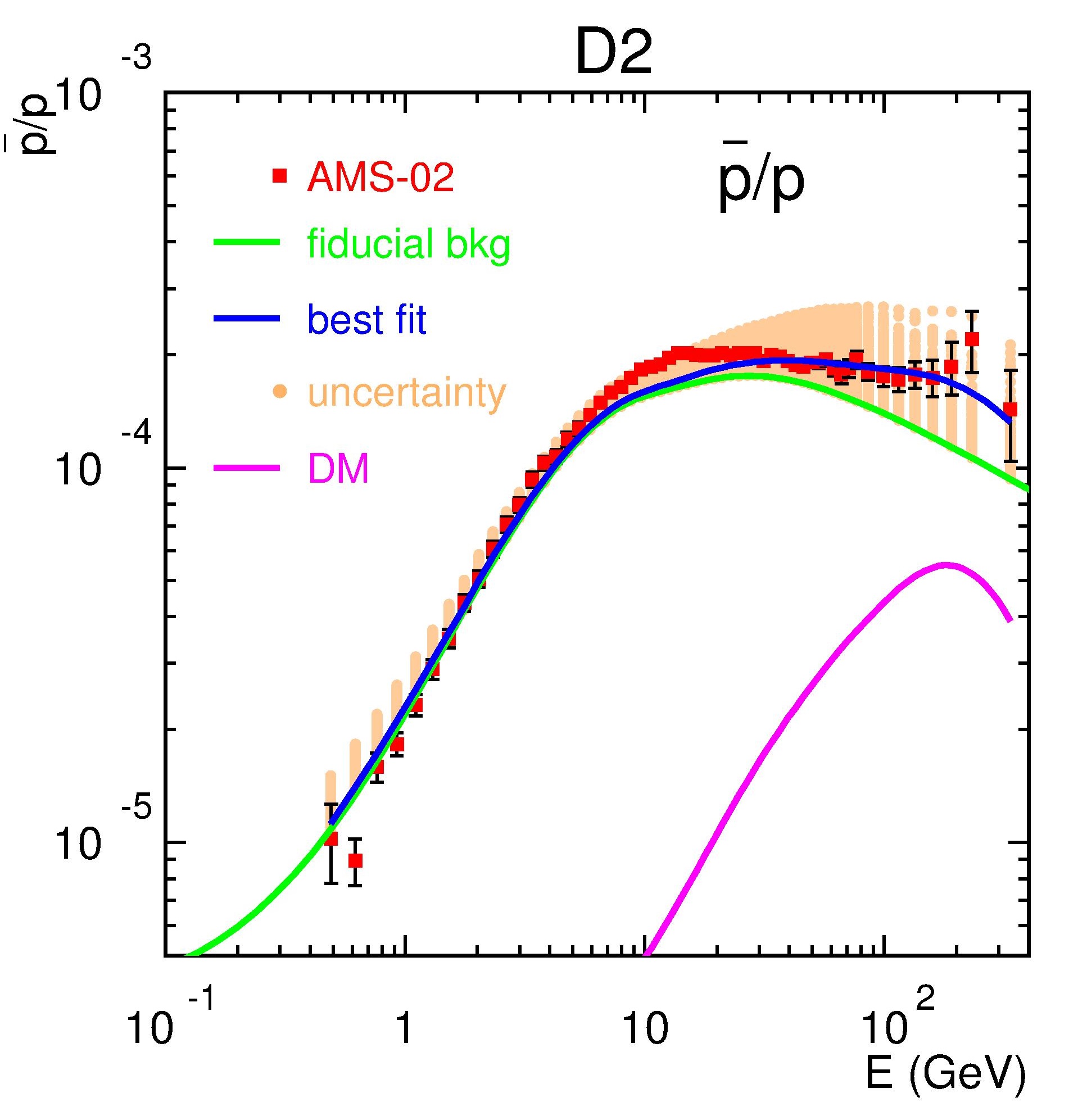}
\end{center}
\caption{Antiproton flux (left) and antiproton-to-proton ratio (right) observed by AMS-02 (red dots and dark error bars) in the simplified dark matter model $D_2$.  The blue solid line shows the prediction of the total cosmic ray flux with dark matter parameter values that best fit the AMS-02 data. The total predicted flux is the sum of the background flux (green curve) and the dark matter contribution (purple curve).  Salmon dots indicate the $2\sigma$ confidence region of the prediction. 
}
\label{fig:fit1}
\end{figure}

\begin{figure}[t]
\begin{center}
\includegraphics[scale=1,width=7cm]{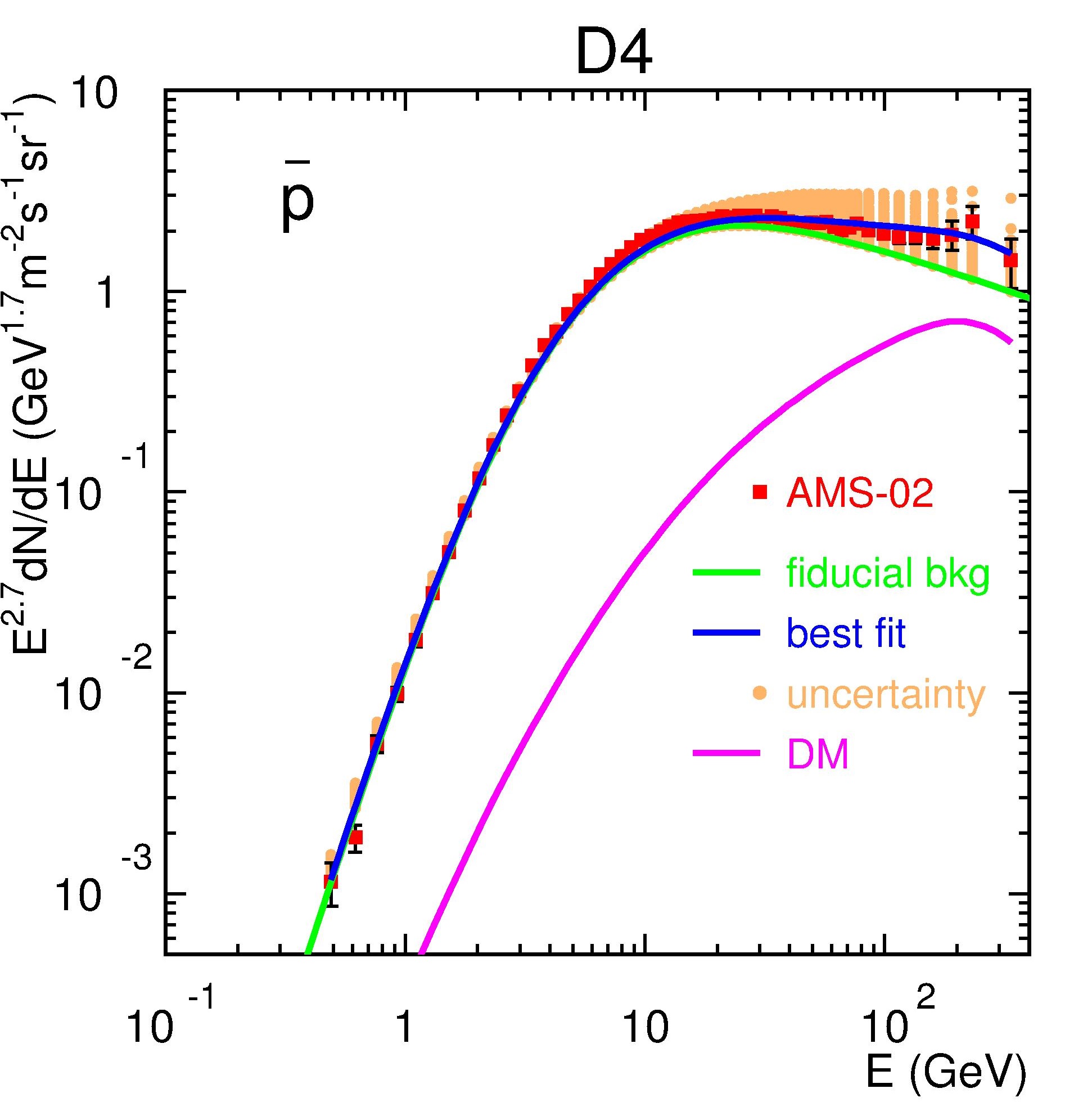}
\includegraphics[scale=1,width=7cm]{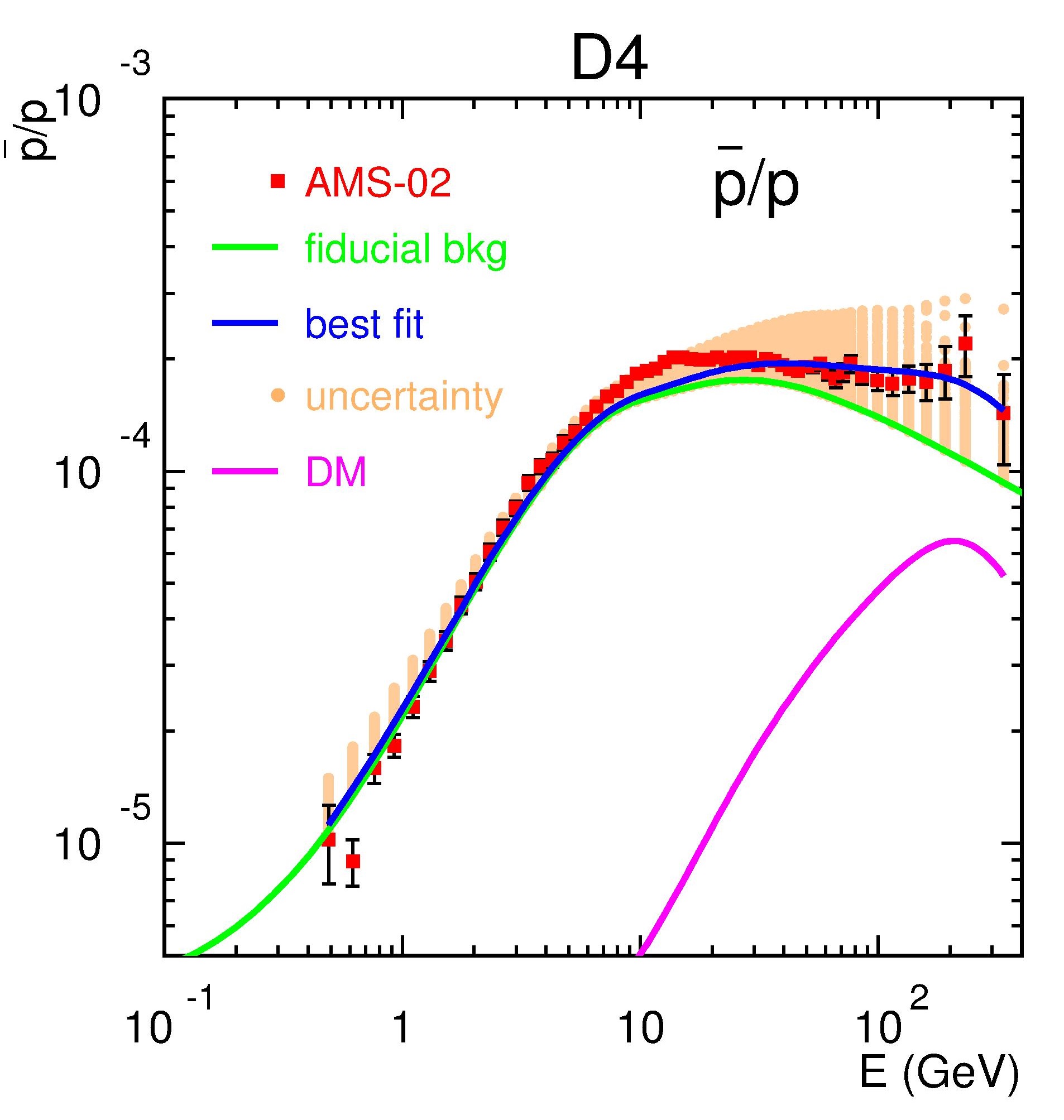}
\end{center}
\caption{Antiproton flux (left) and antiproton-to-proton ratio (right) observed by AMS-02 (red dots and dark error bars) in the simplified dark matter model $D_4$.
}
\label{fig:fit2}
\end{figure}

\begin{figure}[t]
\begin{center}
\includegraphics[scale=1,width=7cm]{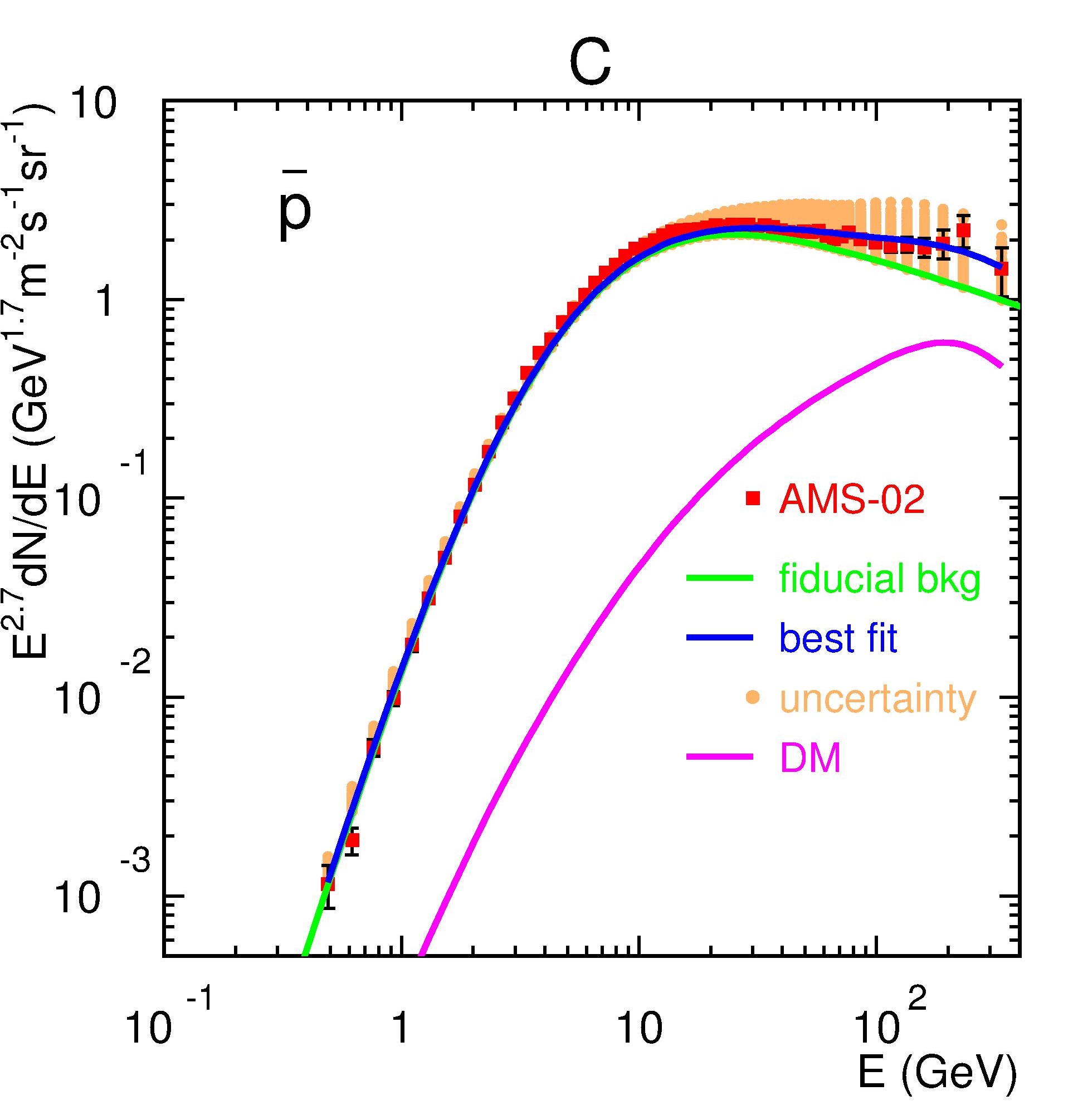}
\includegraphics[scale=1,width=7cm]{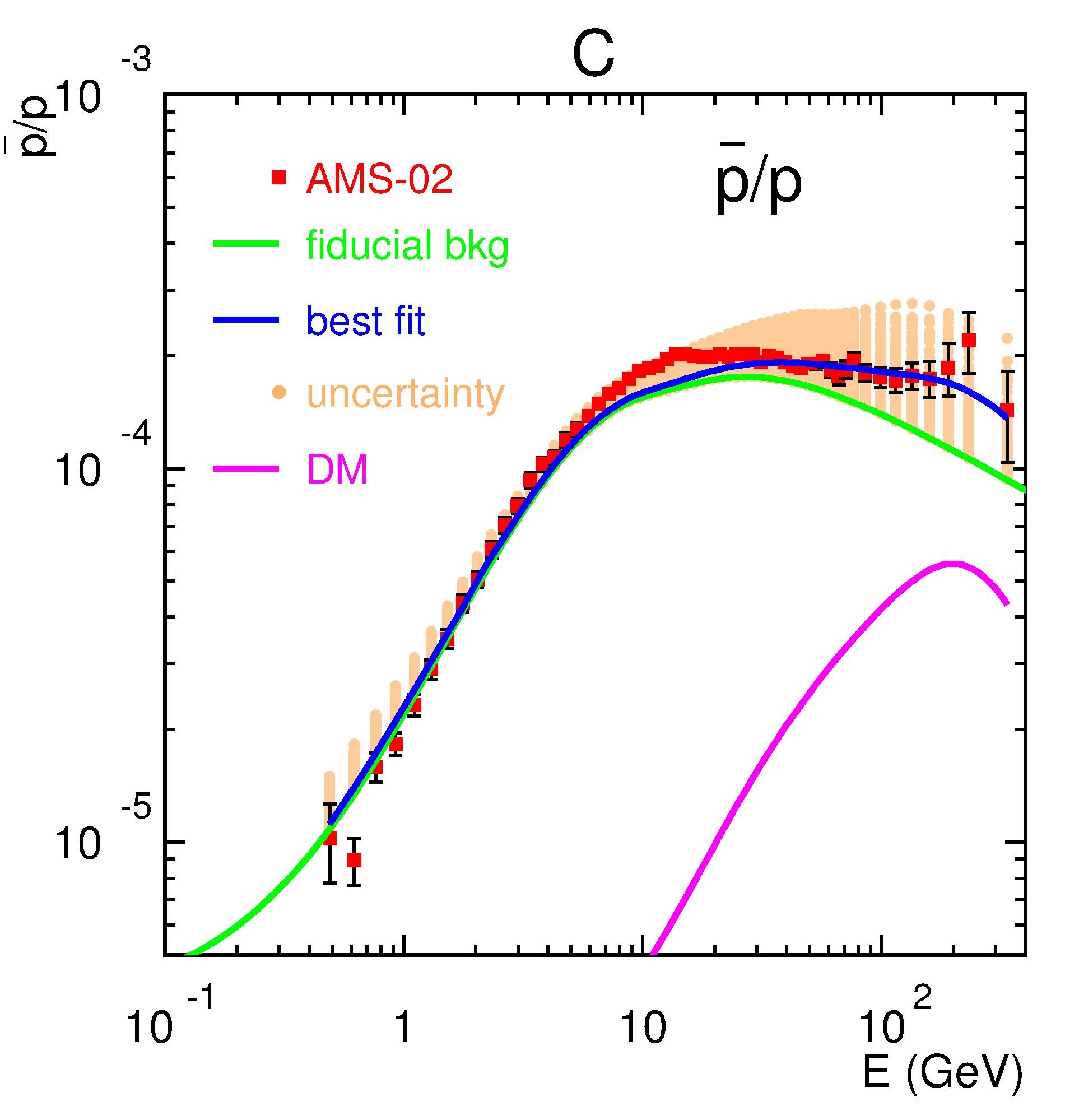}
\end{center}
\caption{Antiproton flux (left) and antiproton-to-proton ratio (right) observed by AMS-02 (red dots and dark error bars) in the simplified dark matter model $C$.
}
\label{fig:fit3}
\end{figure}

\begin{figure}[t]
\begin{center}
\includegraphics[scale=1,width=7cm]{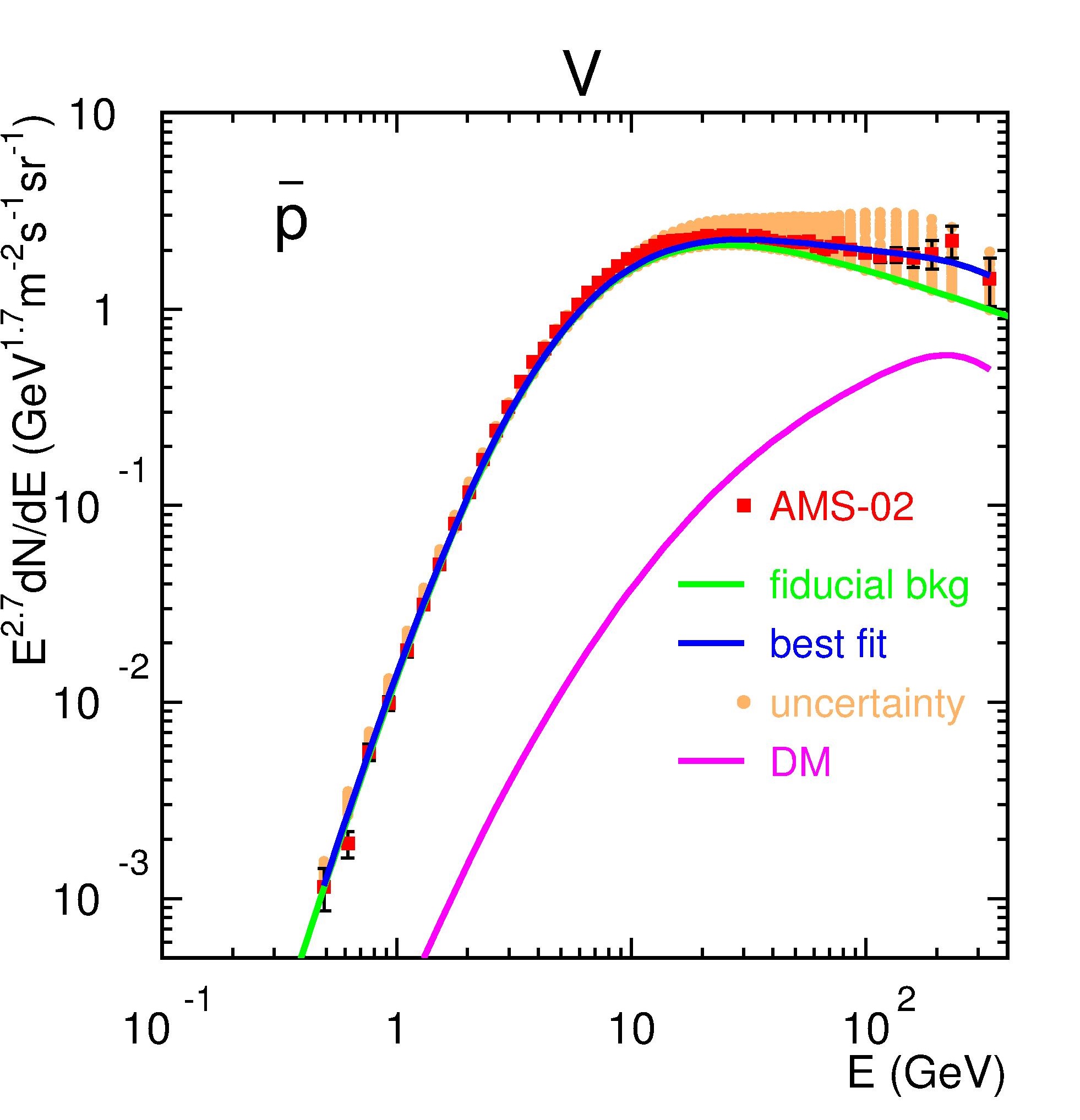}
\includegraphics[scale=1,width=7cm]{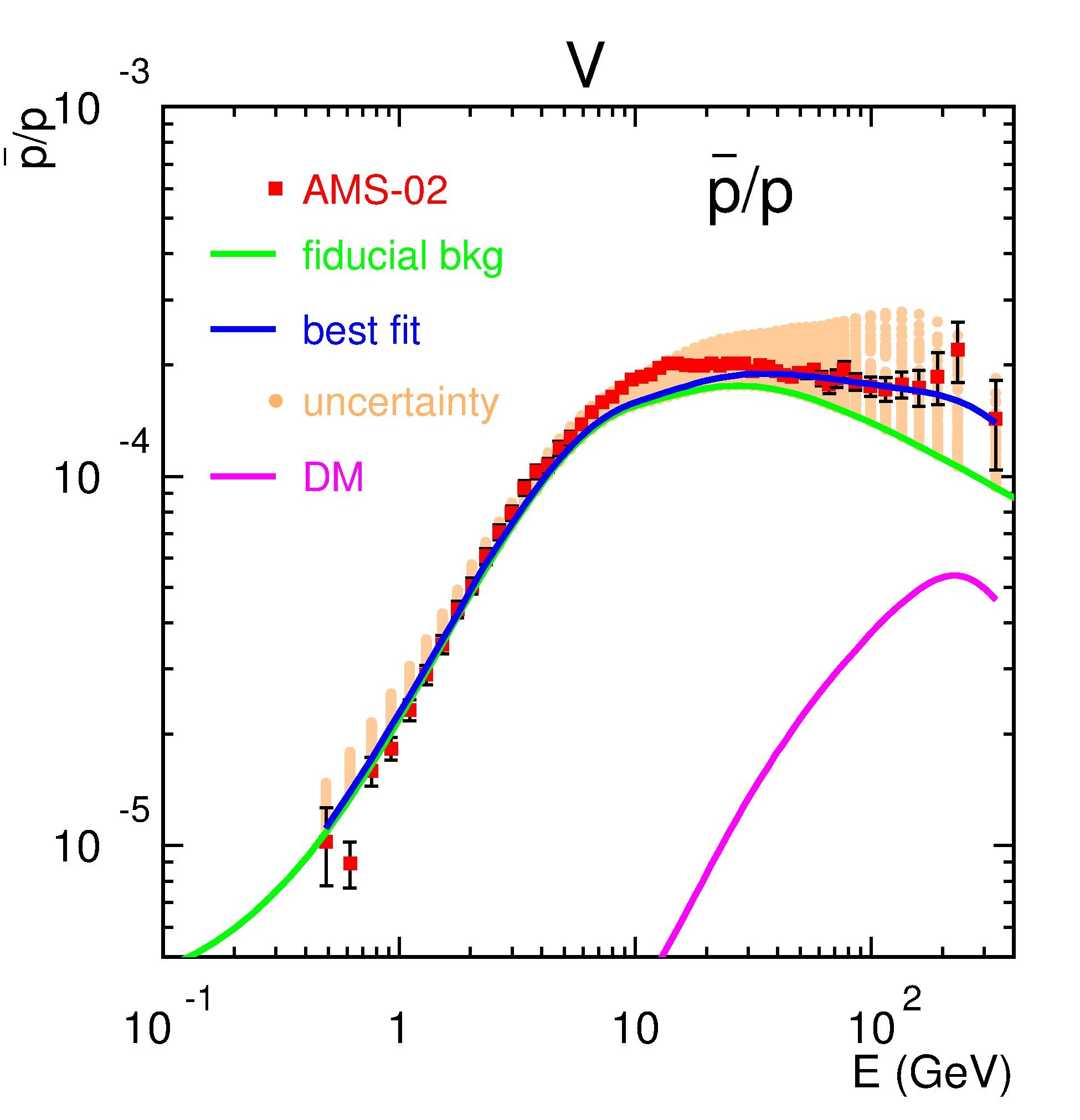}
\end{center}
\caption{Antiproton flux (left) and antiproton-to-proton ratio (right) observed by AMS-02 (red dots and dark error bars) in the simplified dark matter model $V$.
}
\label{fig:fit4}
\end{figure}

In Figs.~\ref{fig:fit1}, \ref{fig:fit2}, \ref{fig:fit3} and \ref{fig:fit4} we show that the AMS-02 antiproton data are consistent with the dark matter framework within the uncertainties for the four models $D_2$, $D_4$, $C$ and $V$ respectively. The left plot in each figure displays the antiproton flux and the right one is the calculated antiproton-to-proton ratio. AMS-02 central values are shown by red dots and the error bars in black indicate experimental uncertainties. The green solid curves are obtained using the parameters shown in Table~\ref{tab:parameter} and display the predicted background originating from standard diffusion process. The dark matter contributions to the observables are then added to the background flux and give the total cosmic ray flux with dark matter contribution that fit the AMS-02 data best (blue solid lines). The dark matter contribution at the best fit point is denoted by purple curve. The combination of the background flux and the dark matter contributions calculated using the parameters in the $2\sigma$ confidence region gives the theoretical uncertainties of the dark matter prediction (salmon colored vertical bars).

Although the most stringent constraints on the simplified models we consider in principle come from indirect detection of dark matter, LHC performed dark matter search using events with large missing transverse momentum plus energetic jet for pseudo-scalar mediator model ($D_4$) at 13 TeV collisions~\cite{CMS-PAS-EXO-16-037}. Their exclusion limit can be directly presented in the plane of dark matter mass vs. mediator mass.
As models $D_2$ and $D_4$ are closely related operators, they should have similar collider constraints and thus we can adopt the LHC constraint on model $D_4$ for $D_2$.

\begin{figure}[t]
\begin{center}
\includegraphics[scale=1,width=7.5cm]{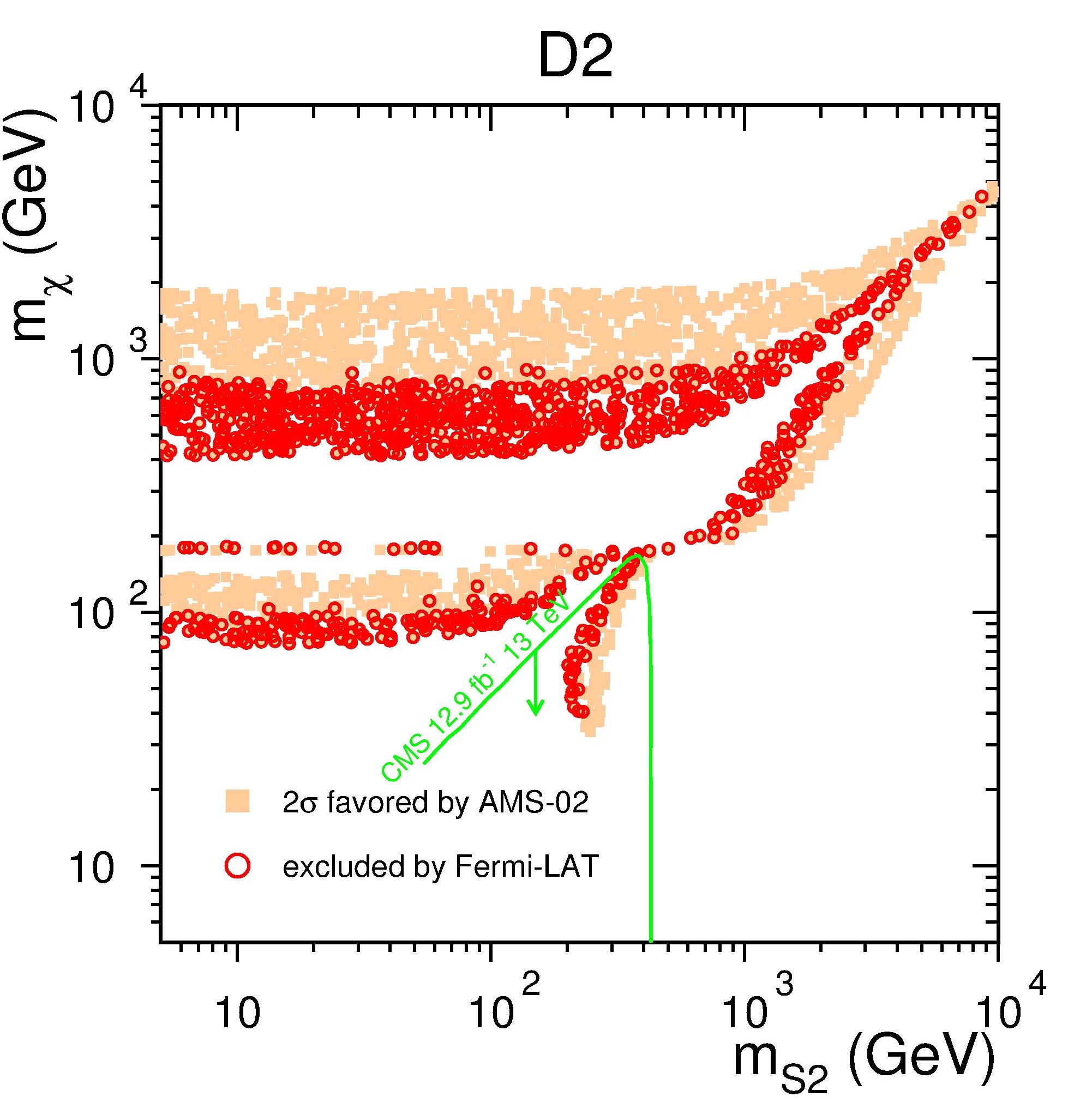}
\includegraphics[scale=1,width=7.5cm]{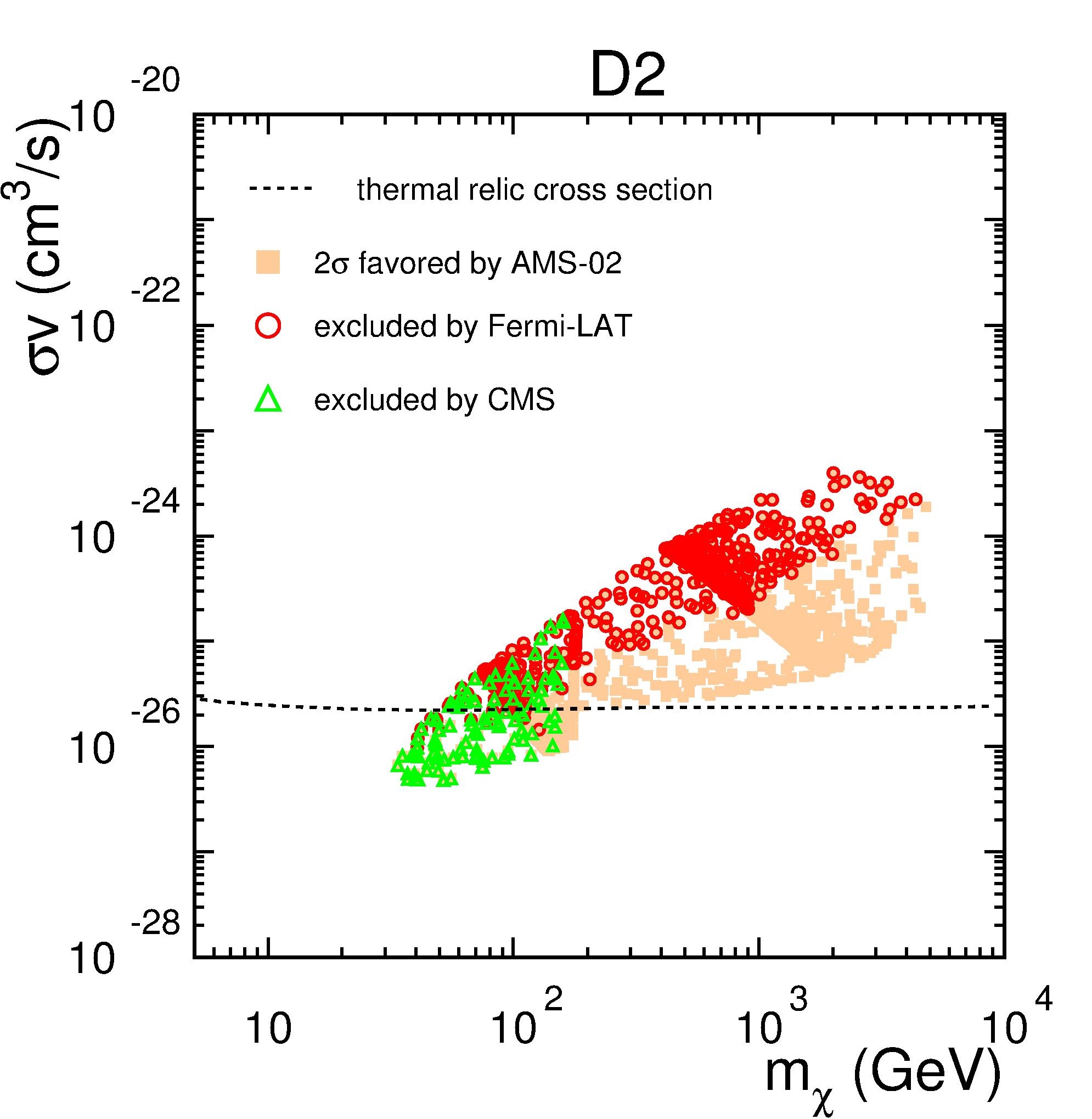}
\end{center}
\caption{Left: the AMS-02 favored region of masses ($m_\chi$ vs. $m_{S_2}$) in the simplified dark matter model $D_2$ we consider. The solid squares estimate $2\sigma$ confidence region. The green curve is the LHC exclusion limit~\cite{CMS-PAS-EXO-16-037}.  Right: the AMS-02 favored region of cross sections ($\sigma v$ vs. $m_\chi$). The green points are excluded by LHC search. The red circles are excluded by Fermi-LAT dSphs. The black dashed curve corresponds to the thermal cross section~\cite{Steigman:2012nb}.}
\label{fig:region1}
\end{figure}

\begin{figure}[t]
\begin{center}
\includegraphics[scale=1,width=7.5cm]{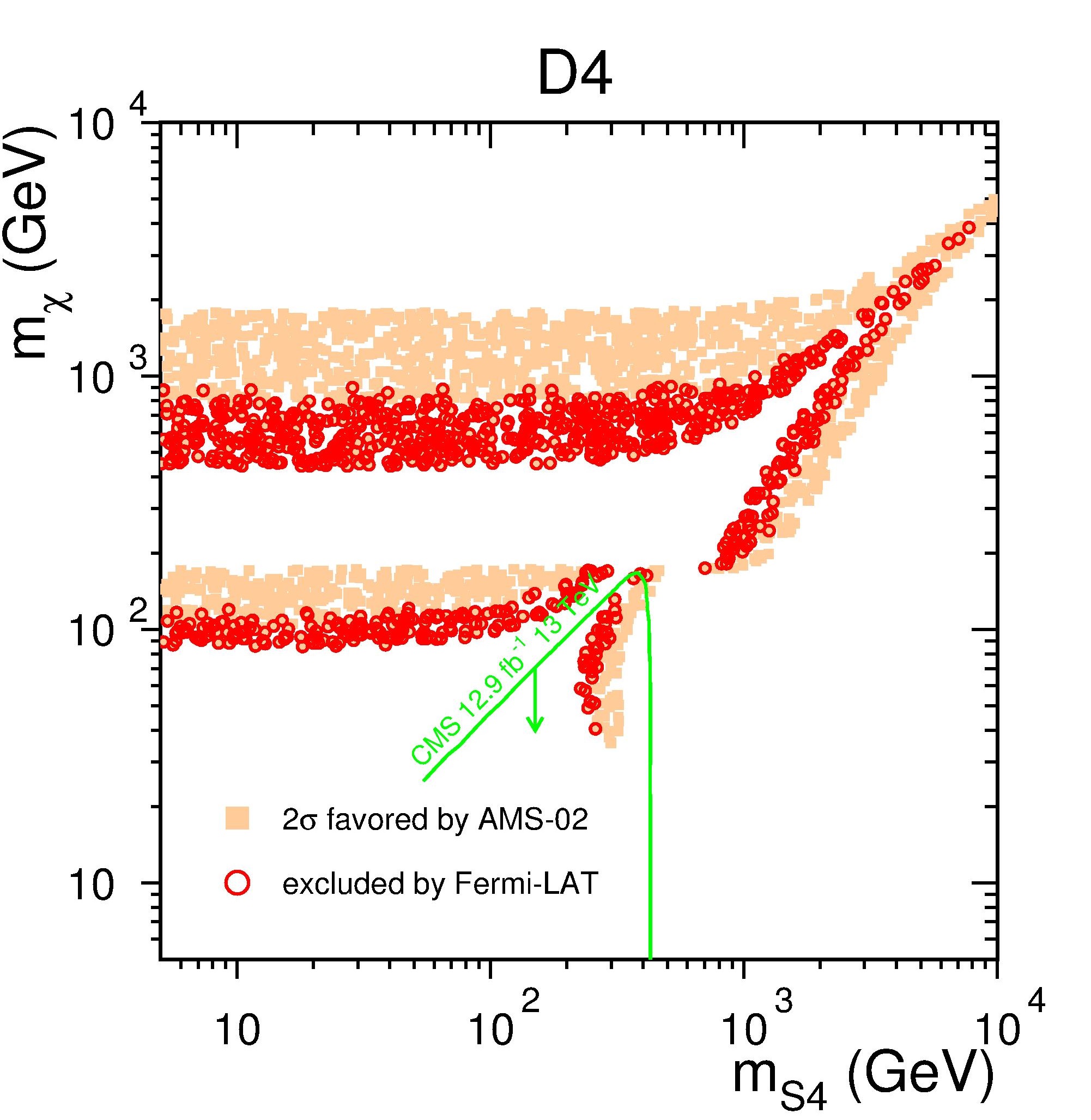}
\includegraphics[scale=1,width=7.5cm]{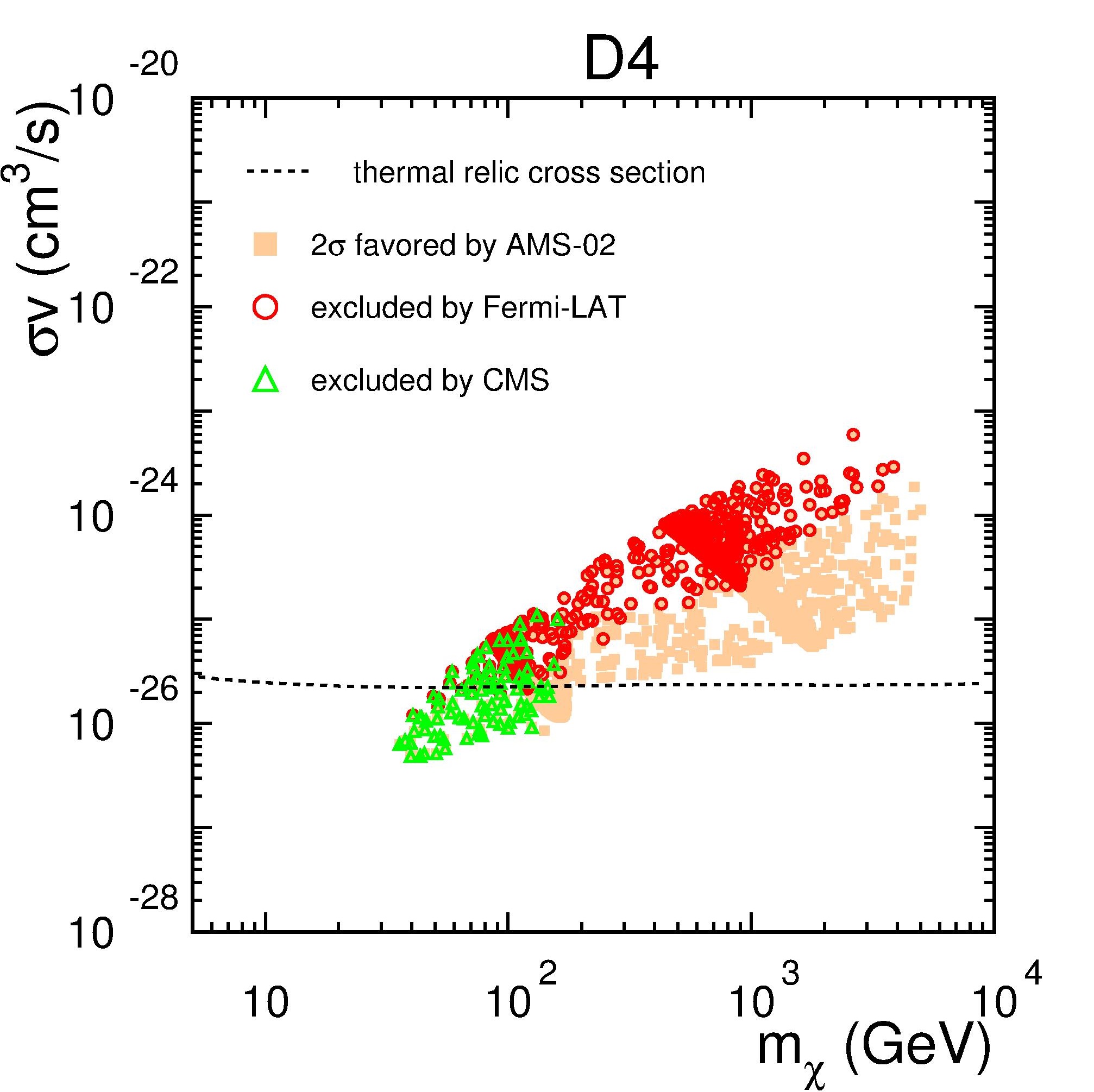}
\end{center}
\caption{Left: the AMS-02 favored region of masses ($m_\chi$ vs. $m_{S_4}$) in the simplified dark matter model $D_4$. Right: the AMS-02 favored region of cross sections ($\sigma v$ vs. $m_\chi$).}
\label{fig:region2}
\end{figure}

\begin{figure}[t]
\begin{center}
\includegraphics[scale=1,width=7.5cm]{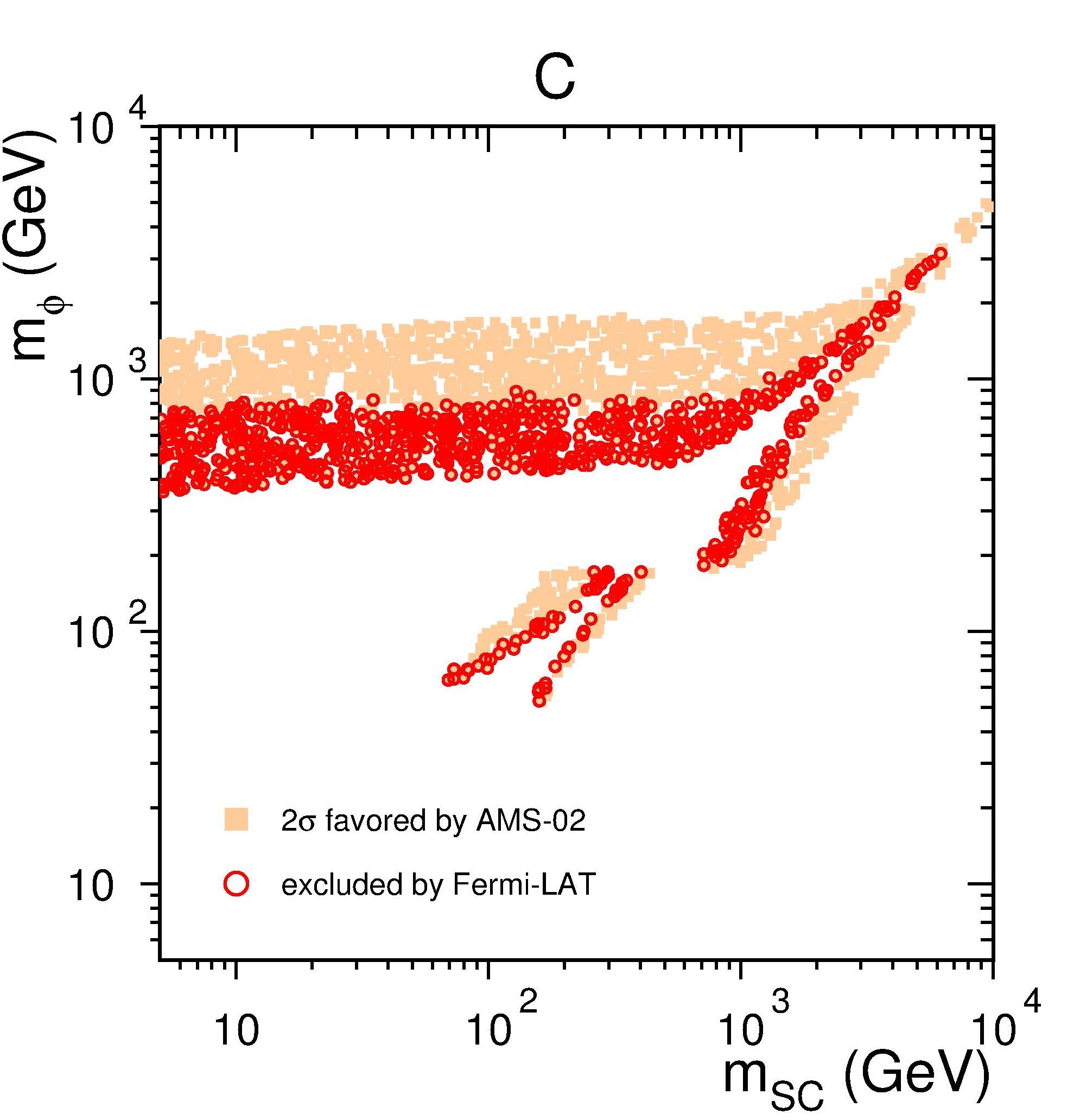}
\includegraphics[scale=1,width=7.5cm]{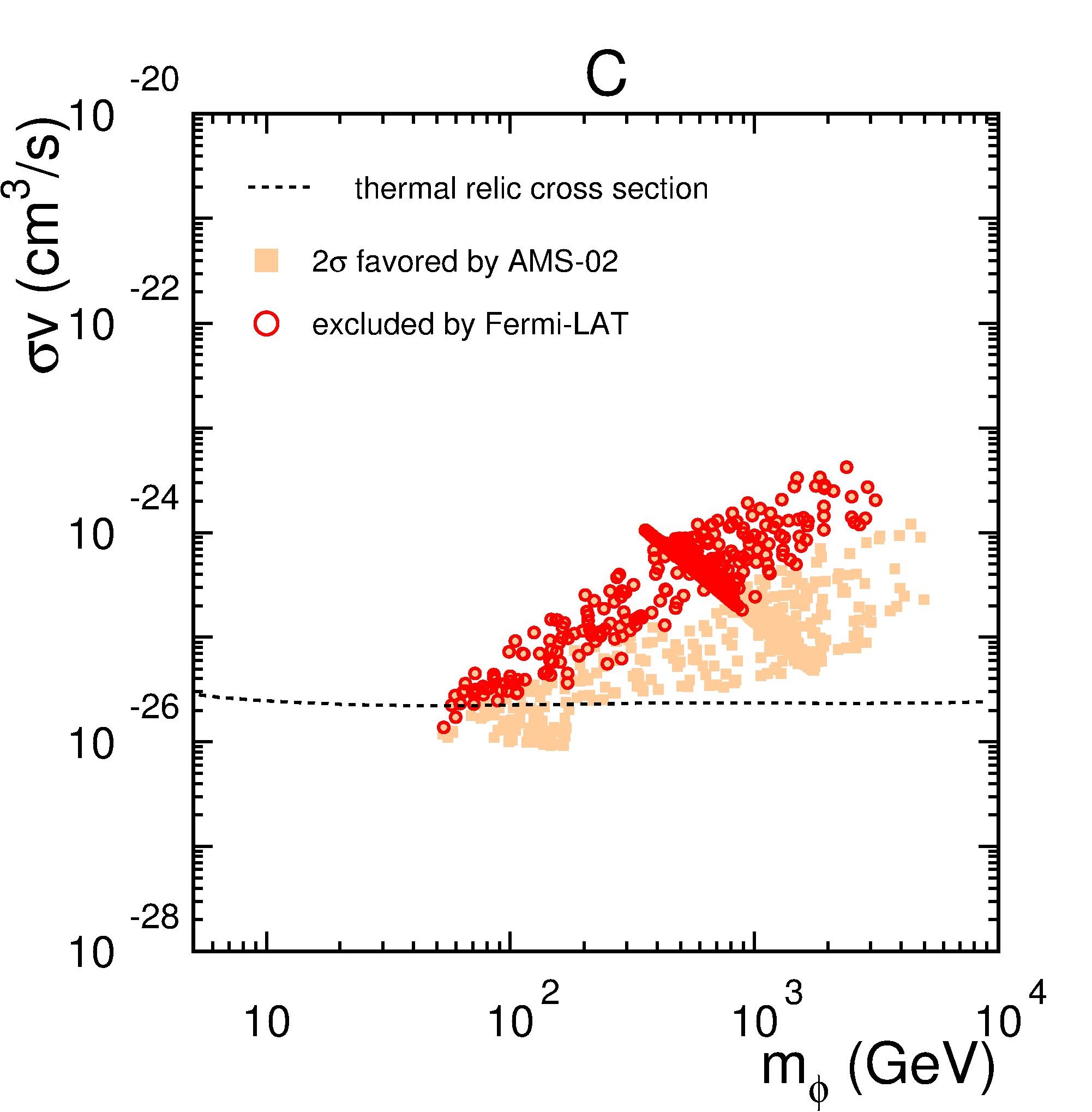}
\end{center}
\caption{Left: the AMS-02 favored region of masses ($m_\phi$ vs. $m_{S_C}$) in the simplified dark matter model $C$. Right: the AMS-02 favored region of cross sections ($\sigma v$ vs. $m_\phi$).}
\label{fig:region3}
\end{figure}

\begin{figure}[t]
\begin{center}
\includegraphics[scale=1,width=7.5cm]{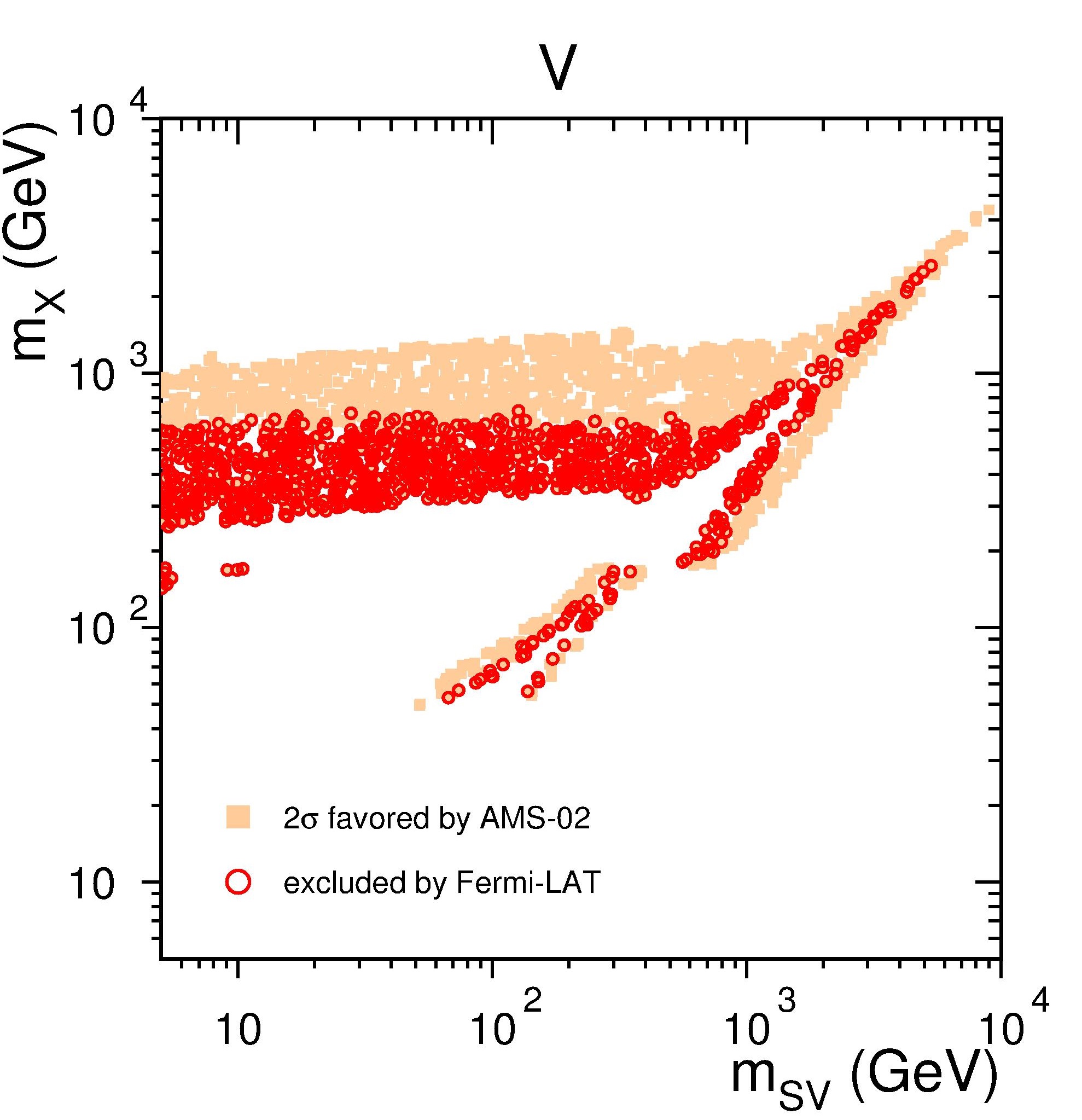}
\includegraphics[scale=1,width=7.5cm]{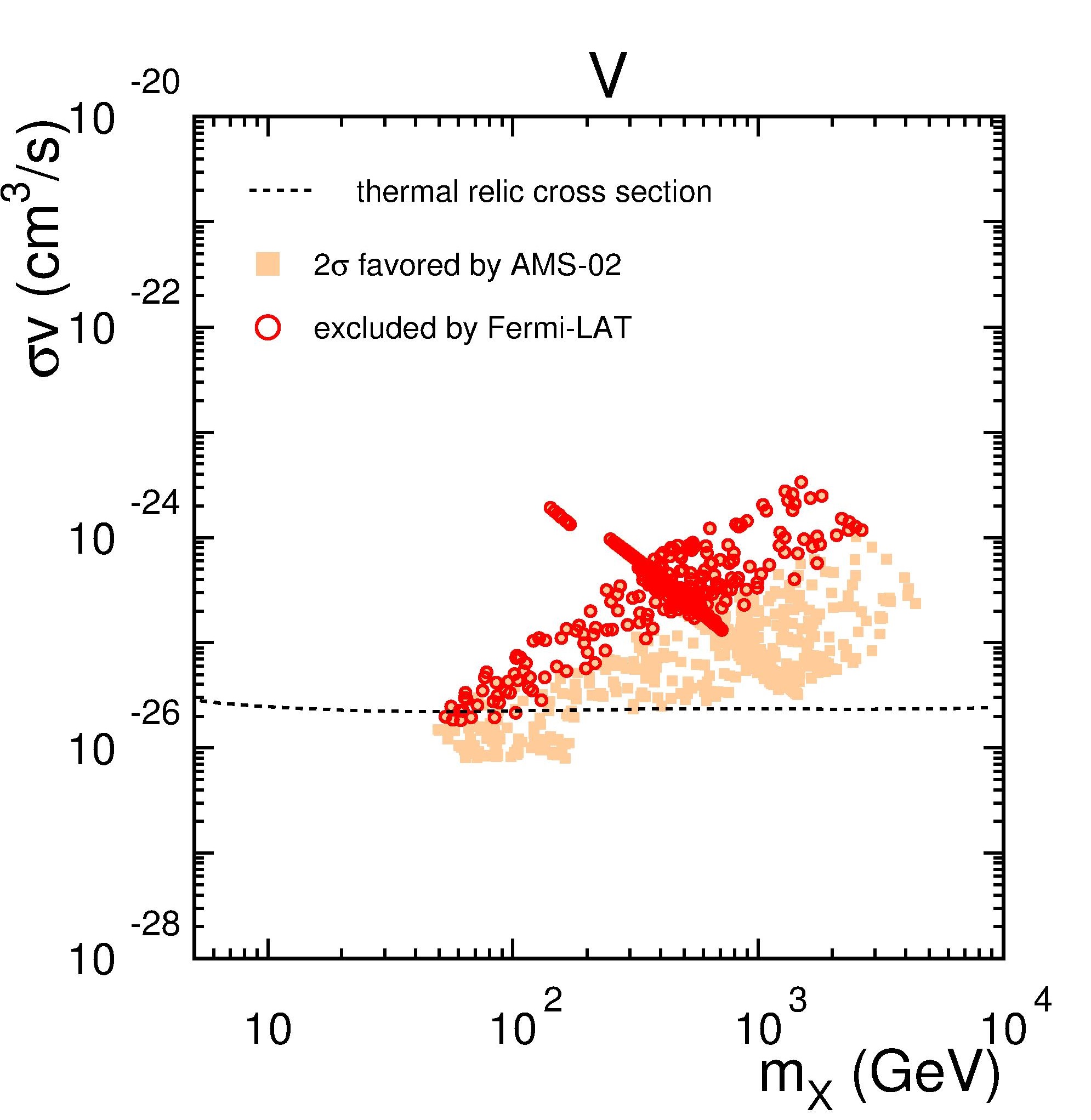}
\end{center}
\caption{Left: the AMS-02 favored region of masses ($m_X$ vs. $m_{S_V}$) in the simplified dark matter model $V$. Right: the AMS-02 favored region of cross sections ($\sigma v$ vs. $m_X$).}
\label{fig:region4}
\end{figure}

In the left frames of Figs.~\ref{fig:region1} and \ref{fig:region2} we show the regions of the mass parameter space preferred by the AMS-02 antiproton data and the constraints from Fermi-LAT dSphs and LHC for models $D_2$ and $D_4$, respectively.  Solid squares denote the estimated $2\sigma$ confidence region. We find that the AMS-02 antiproton data favor $30~{\rm GeV}\lesssim m_\chi\lesssim 5~{\rm TeV}$ region at $2\sigma$ confidence level. The excluded regions by Fermi-LAT dSphs are denoted in red circles. The LHC excludes a small part of the $2\sigma$ confidence region with $m_\chi\lesssim 170$ GeV and $200 \ {\rm GeV}\lesssim m_{S_2}, m_{S_4}\lesssim 420 \ {\rm GeV}$. For scalar and vector dark matter models, the left frames of Figs.~\ref{fig:region3} and \ref{fig:region4} show that the AMS-02 antiproton data favor dark matter masses in the region of 50 GeV - 5 TeV.

The right frames of Figs.~\ref{fig:region1} and \ref{fig:region2} show that the AMS-02 data require an effective dark matter annihilation cross section in the region of $4 \times 10^{-27}$ -- $4 \times 10^{-24}$ ${\rm cm}^3/{\rm s}$ for both $D_2$ and $D_4$ models at about $2\sigma$ C.L. The LHC excludes a part of the low dark matter mass region, denoted by green stars. The relatively small cross section region in red circles can evade the limit from Fermi-LAT dSphs, for instance $\langle \sigma_{\rm ann} v\rangle \lesssim 2 \times 10^{-26} \ {\rm cm}^3/{\rm s}$ allowed for $m_\chi\simeq 100$ GeV and $\langle \sigma_{\rm ann} v\rangle \lesssim 2 \times 10^{-25} \ {\rm cm}^3/{\rm s}$ for $m_\chi\simeq 1$ TeV. The annihilation cross section region $7 \times 10^{-27}$ -- $4 \times 10^{-24}$ ${\rm cm}^3/{\rm s}$ is favored for scalar and vector dark matter as displayed in the right frames of Figs.~\ref{fig:region3} and \ref{fig:region4}.

\section{Conclusions}
\label{sec:Concl}

In this work we investigate the simplified dark matter models favored by the recent AMS-02 antiproton data and consider the constraint from no gamma ray excess in Milky Way dSphs.
The propagation and injection parameters of cosmic rays are determined by fitting the latest AMS-02 data of nuclei fluxes and the secondary antiproton flux is obtained as the fiducial background. In addition to the standard astrophysical cosmic ray, we include a fermion, scalar or vector dark matter component from four simplified models with leptophobic spin-0 mediators that couple only to SM quarks and dark matter particles via scalar and/or pseudo-scalar bilinear. The WIMP-nucleon scattering cross sections and the events with large missing energy plus energetic jet at collider search are both suppressed for the simplified models we consider.

We have shown that the dark matter contribution to the background flux gives a better fit to the data.
%
%
The observation of antiproton prefers dark matter masses in the region of 30 (50) GeV - 5 TeV for simplified fermion (scalar and vector) dark matter models at about $2\sigma$ confidence level. The AMS-02 data also require fermion (scalar and vector) dark matter annihilation cross section as $4 \times 10^{-27} \ (7 \times 10^{-27})$ -- $4 \times 10^{-24}$ ${\rm cm}^3/{\rm s}$.
The LHC excludes a part of the favored region with $m_\chi\lesssim 170$ GeV for fermion dark matter models. The relatively small cross section region can evade the limit from Fermi-LAT dSphs, for instance $\langle \sigma_{\rm ann} v\rangle \lesssim 2 \times 10^{-26} \ {\rm cm}^3/{\rm s}$ for about 100 GeV dark matter mass.

\acknowledgments
We would like to thank Yi Cai for helping with Minuit.
The National Computational Infrastructure (NCI), the Southern Hemisphere's fastest supercomputer, is also gratefully acknowledged.

\appendix
\section{Expressions of mediator decay widths and dark matter annihilation cross sections}
\subsection{fermion dark matter model $D_2$}
The mediator decay widths for the $S_2$ mediator case:
\begin{eqnarray}
\Gamma_{S_2\to \bar{\chi}\chi} &=& {(g_{\chi}^{\rm D2})^2m_{S_2}\over 8\pi} \left(1-{4m_\chi^2\over m_{S_2}^2}\right)^{1/2},\\
\Gamma_{S_2\to \bar{q}q} &=& N_c{(g_q^{\rm D2})^2m_{S_2}\over 8\pi} {m_q^2\over v_0^2} \left(1-{4m_q^2\over m_{S_2}^2}\right)^{3/2} \ \ \ q=u,d,s,c,b,t,\\
\Gamma_{S_2\to gg} &=& {(g_q^{\rm D2})^2\alpha_s^2(m_{S_2})m_{S_2}^3\over 32\pi^3 v_0^2}\left|{4m_t^2\over m_{S_2}^2}\left[1+(1-{4m_t^2\over m_{S_2}^2}){\rm arctan}^2\left(({4m_t^2\over m_{S_2}^2}-1)^{-1/2}\right)\right]\right|^2,\\
\Gamma_{S_2} &=& \Gamma_{S_2\to \bar{\chi}\chi}+ \Gamma_{S_2\to \bar{q}q}+\Gamma_{S_2\to gg}
\end{eqnarray}

The dark matter annihilation cross sections for the $S_2$ mediator case:
\begin{eqnarray}
&&\sigma_{\rm ann} v(\bar{\chi}\chi\to S_2\to \bar{q}q) = {(g_{\chi}^{\rm D2})^2(g_q^{\rm D2})^2N_c\over (4m_\chi^2-m_{S_2}^2)^2+m_{S_2}^2\Gamma_{S_2}^2}{m_\chi^2\over 2\pi} {m_q^2\over v_0^2} \left(1-{m_q^2\over m_\chi^2}\right)^{3/2},\\
&&\sigma_{\rm ann} v(\bar{\chi}\chi\to S_2S_2) = (g_{\chi}^{\rm D2})^4{m_\chi^2(m_\chi^4-2m_\chi^2m_{S_2}^2+m_{S_2}^4)\over 24\pi (2m_\chi^2-m_{S_2}^2)^4}\left(1-{m_{S_2}^2\over m_\chi^2}\right)^{1/2}v^2,
\end{eqnarray}
where $v\simeq 10^{-3}$.

\subsection{fermion dark matter model $D_4$}
The mediator decay widths for the $S_4$ mediator case~\cite{Boveia:2016mrp}:
\begin{eqnarray}
\Gamma_{S_4\to \bar{\chi}\chi} &=& {(g_{\chi}^{\rm D4})^2m_{S_4}\over 8\pi} \left(1-{4m_\chi^2\over m_{S_4}^2}\right)^{1/2},\\
\Gamma_{S_4\to \bar{q}q} &=& N_c{(g_q^{\rm D4})^2m_{S_4}\over 8\pi} {m_q^2\over v_0^2} \left(1-{4m_q^2\over m_{S_4}^2}\right)^{1/2} \ \ \ q=u,d,s,c,b,t,\\
\Gamma_{S_4\to gg} &=& {(g_q^{\rm D4})^2\alpha_s^2(m_{S_4})m_{S_4}^3\over 32\pi^3 v_0^2}\left|{4m_t^2\over m_{S_4}^2}{\rm arctan}^2\left(\left({4m_t^2\over m_{S_4}^2}-1\right)^{-1/2}\right)\right|^2,\\
\Gamma_{S_4} &=& \Gamma_{S_4\to \bar{\chi}\chi}+ \Gamma_{S_4\to \bar{q}q}+\Gamma_{S_4\to gg}
\end{eqnarray}

The dark matter annihilation cross sections for the $S_4$ mediator case~\cite{Arina:2014yna}:
\begin{eqnarray}
&&\sigma_{\rm ann} v(\bar{\chi}\chi\to S_4\to \bar{q}q) = {(g_{\chi}^{\rm D4})^2(g_q^{\rm D4})^2N_c\over (4m_\chi^2-m_{S_4}^2)^2+m_{S_4}^2\Gamma_{S_4}^2}{m_\chi^2\over 2\pi} {m_q^2\over v_0^2} \left(1-{m_q^2\over m_\chi^2}\right)^{1/2},\\
&&\sigma_{\rm ann} v(\bar{\chi}\chi\to S_4S_4) = (g_{\chi}^{\rm D4})^4{m_\chi^2(m_\chi^4-2m_\chi^2m_{S_4}^2+m_{S_4}^4)\over 24\pi (2m_\chi^2-m_{S_4}^2)^4}\left(1-{m_{S_4}^2\over m_\chi^2}\right)^{1/2}v^2,
\end{eqnarray}
where $v\simeq 10^{-3}$.

\subsection{scalar dark matter model $C$}
The $S_C$ mediator decay widths for the scalar dark matter case~\cite{He:2016mls}:
\begin{eqnarray}
\Gamma_{S_C\to \phi^\dagger\phi} &=& {(g_\phi^{\rm C})^2 m_\phi^2\over 16\pi m_{S_C}} \left(1-{4m_\phi^2\over m_{S_C}^2}\right)^{1/2},\\
\Gamma_{S_C\to \bar{q}q} &=& N_c{(g_q^{\rm C})^2m_{S_C}\over 8\pi} {m_q^2\over v_0^2} \left(1-{4m_q^2\over m_{S_C}^2}\right)^{1/2} \ \ \ q=u,d,s,c,b,t,\\
\Gamma_{S_C\to gg} &=& {(g_q^{\rm C})^2\alpha_s^2(m_{S_C})m_{S_C}^3\over 32\pi^3 v_0^2}\left|{4m_t^2\over m_{S_C}^2}{\rm arctan}^2\left(\left({4m_t^2\over m_{S_C}^2}-1\right)^{-1/2}\right)\right|^2,\\
\Gamma_{S_C} &=& \Gamma_{S_C\to \phi^\dagger\phi}+ \Gamma_{S_C\to \bar{q}q}+\Gamma_{S_C\to gg}
\end{eqnarray}

The dark matter annihilation cross sections for the $S_C$ mediator case~\cite{He:2016mls}:
\begin{eqnarray}
&&\sigma_{\rm ann} v(\phi^\dagger\phi\to S_C\to \bar{q}q) = {(g_{\phi}^{\rm C})^2(g_q^{\rm C})^2N_c\over (4m_\phi^2-m_{S_C}^2)^2+m_{S_C}^2\Gamma_{S_C}^2}{m_\phi^2 m_q^2\over 4\pi v_0^2}\left(1-{m_q^2\over m_\phi^2}\right)^{1/2},\\
&&\sigma_{\rm ann} v(\phi^\dagger\phi\to S_CS_C) = {(g_{\phi}^{\rm C})^4 m_\phi (m_\phi^2-m_{S_C}^2)^{1/2}\over 16\pi (2m_\phi^2-m_{S_C}^2)^2}.
\end{eqnarray}

\subsection{vector dark matter model $V$}
The $S_V$ mediator decay widths for the vector dark matter case~\cite{Baek:2014goa}:
\begin{eqnarray}
\Gamma_{S_V\to X^\dagger X} &=& {(g_{X}^{\rm V})^2 m_{S_V}^3\over 64\pi m_X^2}\left(1-{4m_X^2\over m_{S_V}^2}+{12m_X^4\over m_{S_V}^4}\right)\left(1-{4m_X^2\over m_{S_V}^2}\right)^{1/2},\\
\Gamma_{S_V\to \bar{q}q} &=& N_c{(g_q^{\rm V})^2m_{S_V}\over 8\pi} {m_q^2\over v_0^2} \left(1-{4m_q^2\over m_{S_V}^2}\right)^{1/2} \ \ \ q=u,d,s,c,b,t,\\
\Gamma_{S_V\to gg} &=& {(g_q^{\rm V})^2\alpha_s^2(m_{S_V})m_{S_V}^3\over 32\pi^3 v_0^2}\left|{4m_t^2\over m_{S_V}^2}{\rm arctan}^2\left(\left({4m_t^2\over m_{S_4}^2}-1\right)^{-1/2}\right)\right|^2,\\
\Gamma_{S_V} &=& \Gamma_{S_V\to X^\dagger X}+ \Gamma_{S_V\to \bar{q}q}+\Gamma_{S_V\to gg}
\end{eqnarray}

The dark matter annihilation cross sections for the $S_V$ mediator case~\cite{Baek:2014goa,Arcadi:2017kky}:
\begin{eqnarray}
&&\sigma_{\rm ann} v(X^\dagger X\to S_V\to \bar{q}q) = {(g_{X}^{\rm V})^2(g_q^{\rm V})^2N_c\over (4m_X^2-m_{S_V}^2)^2+m_{S_V}^2\Gamma_{S_V}^2}{m_X^2 m_q^2\over 12\pi v_0^2} \left(1-{m_q^2\over m_X^2}\right)^{1/2},\\
&&\sigma_{\rm ann} v(X^\dagger X\to S_VS_V) = {(g_X^{\rm V})^4 \over 144\pi m_X^2}{6m_X^4-4m_X^2m_{S_V}^2+m_{S_V}^4\over (2m_X^2-m_{S_V}^2)^2}\left(1-{m_{S_V}^2\over m_X^2}\right)^{1/2}.
\end{eqnarray}


\end{document}